# Molecularly imprinted nanopores for multiplexed sensing, release, and in-edge computing

*Ali Douaki\*, Shukun Weng, Silvia Dante, Nako Nakatsuka, Makusu Tsutsui, Roman Krahne, and Denis Garoli\**

A. Douaki, S. Weng, S. Dante, G. Lanzavecchia, A. Sapunova, R. Krahne, D. Garoli

Istituto Italiano di Tecnologia, Via Morego 30, Genova 16163, Italy

N. Nakatsuka

Laboratory of Chemical Nanotechnology, EPFL, Genève, 1202, Switzerland

M. Tsutsui

The Institute of Scientific and Industrial Research, Osaka University, Mihogaoka 8-1Ibaraki Osaka 567-0047, Japan

A. Douaki, G. Lanzavecchia, D. Garoli

Dipartimento di scienze e metodi dell'ingegneria Università di Modena e Reggio Emilia Via Amendola 2, Reggio Emilia 42122, Italy

E-mail: ali.douaki@unimore.it; denis.garoli@unimore.it

# Abstract

In nanopore technology, the development of multiplexed detection and release platforms with high spatial and temporal resolution remains a significant challenge due to the difficulty in distinguishing signals originating from different nanopores in a single chip. In this work, we present a solid-state nanopore system functionalized with molecularly imprinted polymers (MIPs) for the selective detection and controlled release of neurotransmitters. We designed a nanopore array where each nanopore is functionalized with a specific MIP able to recognize specific neurotransmitters (dopamine, gamma-aminobutyric acid, and histamine, respectively). The platform demonstrated high performance in terms of sensitivity, selectivity, recovery, and stability. Multiplexed detection with high spatiotemporal resolution of the order of 100 ms/ 3 µm was achieved by specifically depositing MIPs and conductive hydrogels on different nanopores prepared on a single solid-state membrane. The employment of micro-chambers for each nanopore prevented signal cross-talk, thereby enabling simultaneous detection and release of multiple neurotransmitters. Moreover, we demonstrated computing with different logic gates and in-edge computing. This nanopore platform represents a radically novel approach towards hybrid solid-state nanopores able to perform real-time label-free multiplex detection, controlled biomolecule release, and ionic logic computing, addressing key challenges in neurochemical sensing and bio-computation.

# Introduction

Real-time, multiplexed detection and precise release/delivery of biomolecules are critical for advancing our understanding in biological processes[1–5]. For instance, neurotransmitters such as dopamine, gamma-aminobutyric acid (GABA), and histamine play a key role in orchestrating brain functions, such as cognition, mood, and motor control. Dysregulation of these neurotransmitters is often linked to a range of disorders, including depression, Parkinson's disease, and schizophrenia [1,6–12]. Multiplexed sensing of neurotransmitters can be crucial for understanding complex neural processes such as synaptic plasticity, learning, and memory[13–15]. Many of these biological processes take place on a small spatial scale where the release and uptake of biomolecules occur within highly confined spaces and over very short timescales, typically on the order of milliseconds to tens of milliseconds [13,14,16]. Thus, capturing biological activity in detail requires the development of detection platforms with high spatial and temporal resolution that approach biological processes. This capability is essential for resolving individual events within densely interconnected neural networks. Achieving these two goals is crucial for accurately investigating neural processes in particular, and biological processes more generally [17].

Traditional detection methods based on electrochemical and optical approaches have been extensively used in biosensing [18–21]. However, these methods still suffer either in terms of spatial/temporal resolution, working complexity, and selectivity [4,5,17,18,22–24]. These factors hinder their application in the simultaneous detection of multiple analytes (such as neurotransmitters). In this context, solid-state nanopores are emerging as one of the most promising technologies due to their high sensitivity, label-free detection, real-time monitoring, shape structure control, and high spatial/temporal resolution [19,20,24–28]. Although substantial progresses have been made over the past two decades [29,30], a key challenge remains in achieving multiplexed label free detection of different target molecules using nanopore technologies. Specifically, distinguishing signals from multiple nanopores within a multiplexed system -while maintaining high spatial resolution- is difficult due to signal cross-talk, which hampers measurement and control accuracy.

Molecular imprinting has recently demonstrated significant potential in enhancing the selectivity of biosensors, including electrochemical sensors and field-effect transistors [31–37]. Molecularly imprinted polymers (MIPs) are created by polymerizing functional monomers around a target analyte, which serves as a "template." After polymerization, the template is removed, leaving behind binding cavities that are highly selective for the target molecule [31–33]. When integrated with nanopores, these selective cavities can enable the specific recognition and binding of neurotransmitters [31,37–41]. Despite this potential, the combination of MIPs with nanopores remains largely unexplored, although such an integrated platform could offer real-time, multiplexed detection with high spatial and temporal resolution.

In addition, in-memory and in-edge computing has emerged as an alternative to traditional von Neumann architectures enabling computational capabilities within memory devices themselves, overcoming the limitation in the conventional separation between processing and memory units. This approach has been successfully implemented across various platforms, including resistive devices, photonic systems, crossbar arrays, spin-transfer torque magnetic memories, and memristors [42–46]. Recent developments have begun exploring the integration of sensing capabilities with computation, particularly in ion-sensitive field-effect transistor arrays. These arrays demonstrate dual functionality by sensing ion concentrations and storing calibration data [47]. However, the full experimental integration of sensing and computing within a single memory cell remains largely unexplored. Our work introduces a completely novel concept that we term 'ionic edge-computing/in-memory sensing,' which unifies sensing, delivery and computing capabilities within individual memory cells.

Here, we report, for the first time, the use of MIP-integrated nanopores for real-time, multiplexed sensing and controlled release of neurotransmitters. By integrating MIPs into solid-state nanopores, a highly sensitive and selective platform capable of simultaneously differentiating between multiple neurotransmitters (dopamine, GABA, and histamine) is demonstrated. Furthermore, we employ these MIP nanopores to perform ionic logic computing with different logic gates (NAND, NOR, and NOT), and integrate these platforms into a unified in-edge computing platform capable of detecting neurotransmitters, executing logic computations, and triggering the controlled release of molecules.

# Results and discussion

**Working mechanism of the molecularly imprinted nanopore platform**

Figure 1 illustrates the operating principle of our in-edge computing platform for simultaneous sensing and targeted delivery. The system consists of three key components. First, a multiplexed sensing unit comprising three solid-state nanopores, each functionalized with a different MIP for selective detection of dopamine (DA), GABA, and histamine (His). The MIPs provide highly selective molecular recognition *via* shape-complementary binding, which modulates ionic conductance and serves as a reliable sensing mechanism. To enable multiplexing, we engineered a novel architecture in which spatially separated MIP-functionalized nanopores are integrated on the same membrane and interfaced with selectively deposited conductive hydrogels (Fig. 1a). This design allows for independent current measurements from each nanopore, thus effectively preventing signal cross-talk. All MIPs share a common *cis* chamber (sensing volume) while maintaining discrete *trans* chambers for each nanopore, enabling simultaneous detection of multiple neurotransmitters with high spatial resolution (few μm), addressing a key limitation of traditional nanopore systems.

To perform in-edge computing, the realization of neurotransmitter logic gates is fundamental. Hence, ionic logic gates using MIP nanopores were fabricated, where neurotransmitter presence is translated into binary inputs (presence = 1, absence = 0), allowing fundamental logic gates (NOT, NOR, NAND) to be implemented based on ionic current thresholds (Fig. 1b). These logic functions are executed directly within the sensing unit, therefore eliminating the need for external processors and reducing both latency and energy consumption and allowing for in-edge decision making and triggering (Fig. 1c). Finally, the same platform enables stimulus-triggered release of neurotransmitters *via* the nanopores used for sensing. Upon application of an external voltage, the non-covalent interactions between the trapped neurotransmitters and their MIP binding sites are disrupted, facilitating precise and reversible release of the trapped molecules (Fig. 1d).

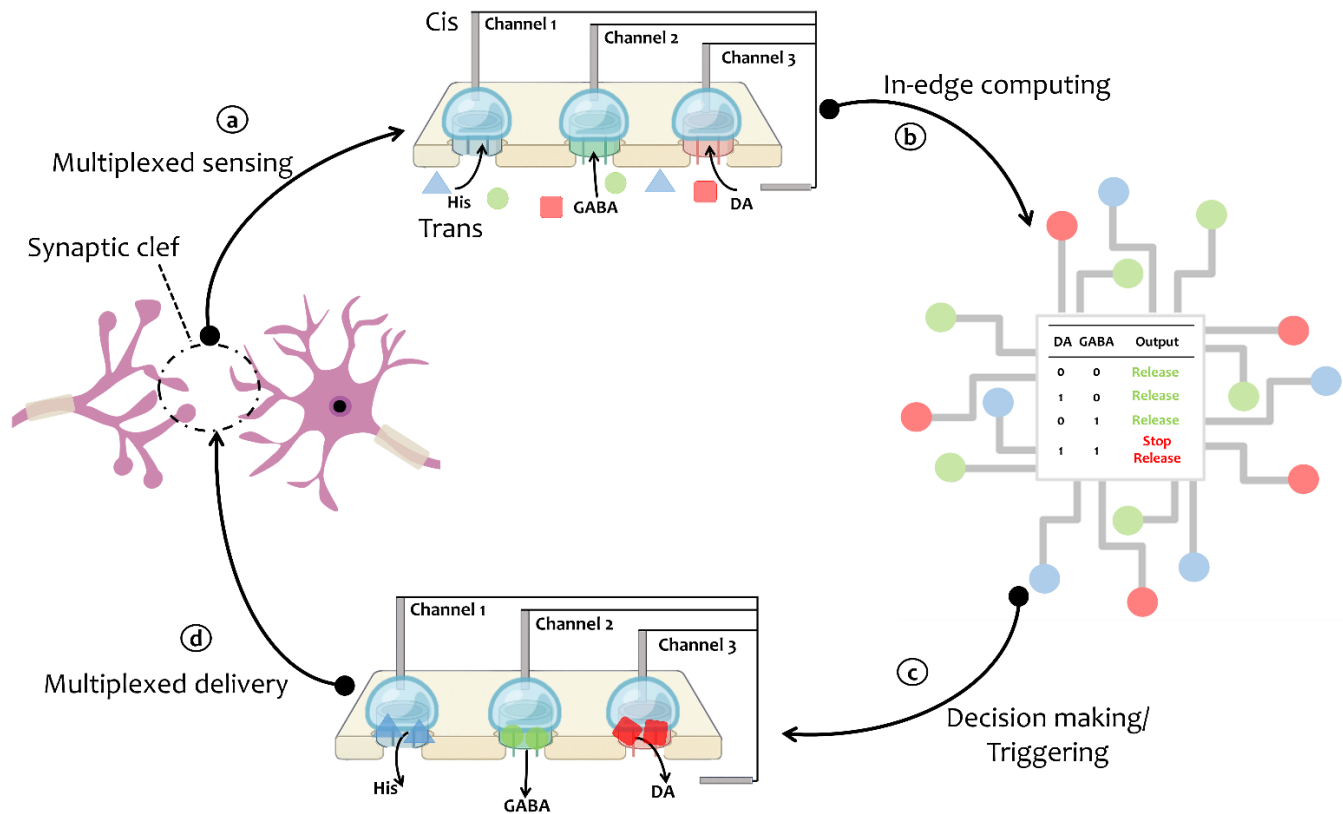

**Figure 1.** Conceptual schematic of a multiplexed solid-state MIP-nanopore platform for neurotransmitter multiplex sensing, logic processing, and multiplexed controlled release. (a) The platform integrates three distinct nanopores functionalized with MIPs for dopamine, GABA, and histamine, enabling specific detection of each neurotransmitter. (b) The signals generated from neurotransmitter binding are employed as inputs for performing logic processing (in-memory computing) where binary operations (e.g., NOR, NAND) are performed based on input patterns. (c-d) Processed outputs from the logic gates then trigger selective release of the desired neurotransmitters, facilitating closed-loop neuromodulation and intelligent bio-interfacing.

## Characterization of single MIP-nanopores

Before demonstrating the multiplexing capabilities of the proposed device, each MIP nanopore was characterized individually. For the dopamine nanopore sensor, we first fabricated a single nanopore on a $Si_3N_4$ membrane using focused ion beam (FIB) milling. This nanopore was then locally functionalized with a dopamine-molecularly imprinted polymer (DA-MIP), which was optimized for selective binding towards dopamine (Fig. 2a). Initially, the deposited MIP still contained the dopamine template, which blocks the passage of ions through the nanopore. Therefore, the template was subsequently removed using ethanol/acetic acid solution. [31,31,48] After template removal, the MIP's morphology was investigated by SEM and AFM, as shown in Fig. S1 and S2 (Supporting Information - SI), which revealed the porous structure of the MIP, featuring cavities that allow ions in the electrolyte to pass through and the target analyte to bind. [33,41]

The MIP nanopore platform presented in this work operates by applying a transmembrane voltage across the $Si_3N_4$ membrane, generating a measurable ionic current. When the target analyte is absent, the MIP cavities remain unoccupied, allowing uninhibited ionic flow through the nanopores (Figure. 2b). When a positive bias is applied to the cis chamber relative to the trans (shared analyte reservoir), the positively charged neurotransmitter molecules are electrophoretically driven from the trans chamber toward the cis chamber under the applied field, enabling binding within the MIP cavities and subsequent modulation of ionic current, a working mechanism similar to aptamer-based nanopore sensors [25,49] Upon introduction of the target analyte, specific binding occurs within the MIP recognition sites, resulting in partial nanopore occlusion and a corresponding reduction in ionic current, which serves as the detection signal. Figure. 2c shows the conductance of the nanopore before and after the removal of the dopamine: as the dopamine template was removed, we observed an increase in the conductance, proving the successful removal. To confirm successful target extraction, ultraviolet–visible (UV–Vis) spectroscopy was performed on the extraction solution before and after the process. A notable decrease in the absorbance peak around 280 nm, which corresponds to the characteristic absorption of dopamine, indicated effective removal of the target molecule (Figure S3 - SI).[33,50]

The sensitivity of the dopamine MIP sensor was assessed by exposing the system to a series of dopamine concentrations and generating a calibration curve. Figures S4 and S5 (SI) illustrate the experimental setup, including both the microfluidic and the electrical measurement apparatus. [28] The microfluidic device is composed of a solid-state MIP-functionalized nanopore chip mounted between two fluidic compartments, designated as the top and bottom chambers. The nanopore, located at the interface of the two chambers, acted as a selective ionic transport channel modulated by dopamine binding within the MIP layer. During the measurements, the top chamber was sequentially filled with dopamine solutions of increasing concentrations, while the bottom chamber contained a buffer to maintain a stable ionic environment across the pore. Changes in nanopore conductance were recorded in response to each concentration step, enabling quantitative analysis of the sensor's sensitivity and limit of detection (LOD). The MIP nanopores were sensitive to dopamine ranging from 1 pM to 100 µM with a LOD of 2.82 pM and a limit of quantification (LOQ) of 23.20 pM for dopamine, aligning well with physiologically relevant concentrations (Fig. 2d and S6) [51,52].

To demonstrate the viability of our sensor in undiluted complex media, the MIP-nanopore performance was also deployed in neurobasal media, as this biofluid lacks endogenous dopamine but contains a variety of nonspecific amino acids and proteins essential for supporting neuronal cultures *in vitro* (Figure. 2e and S7). The selectivity of a biosensor is critical in complex, real-world environments where various interfering molecules are present. [52,53] Thus, measurements were conducted in neurobasal medium spiked with different structurally similar neurotransmitters to the specific targets of interest, including norepinephrine (NE), a monoamine "serotonin" (5-HT), GABA and histamine. These molecules were chosen to test the selectivity against the other targets.[18,30,54] As can be seen from Figure 2f and Figure S8 (SI), NE and serotonin showed negligible effects on the ionic current, with only a 5.75 ± 0.56% and 3.75 ± 0.2% conductance reduction, respectively, at concentrations of 1 nM. This selectivity is attributed to the high binding affinity of the DA-MIP cavities for dopamine, which exclude NE and 5-HT due to subtle differences in molecular structure. The cross-selectivity experiments further confirmed that dopamine binding induced a significantly larger current reduction (33.03 ±0.47 %) compared to NE and 5-HT, demonstrating the excellent specificity of the sensor, as was previously reported [31]. Such selectivity in the presence of interferents renders MIP-based nanopores advantageous compared to traditional sensors that suffer from cross reactivity, and voltametric methods that have difficulties in distinguishing dopamine analogs with overlapping oxidation signals [29].

To evaluate the reusability and operational stability of the DA-MIP-based sensor for long-term dopamine detection, the sensor was subjected to six consecutive sensing-regeneration cycles. Each cycle involved exposure to 1 µM dopamine, followed by electrochemical regeneration achieved by reversing the applied potential to facilitate desorption of the bound dopamine molecules from the MIP cavities. Following each regeneration step, the sensor's response to 1 µM dopamine was reassessed. The sensor retained over 95% of its initial sensitivity across all cycles (Fig. 2g), indicating reusability with minimal loss of recognition capabilities. Additionally, the long-term stability of the sensor was evaluated by measuring its response to 1 µM dopamine at three-day intervals over a period of nine days. The sensor exhibited a consistent current suppression of 97.2 ± 7.0 %, suggesting negligible degradation of the MIP layer and preservation of molecular recognition over time (Fig. 2h). These results collectively demonstrate the robustness and suitability of the DA-MIP sensor for repeated and prolonged sensing applications. Similarly, GABA-specific and histamine-specific MIP-nanopores were also fabricated, addressing fundamental neurotransmitters involved in various brain functionalities. [38,55–57] These MIP-nanopores were characterized by a similar set of experiments as the dopamine-MIP sensor (please refer to the Notes 2-3 in SI).

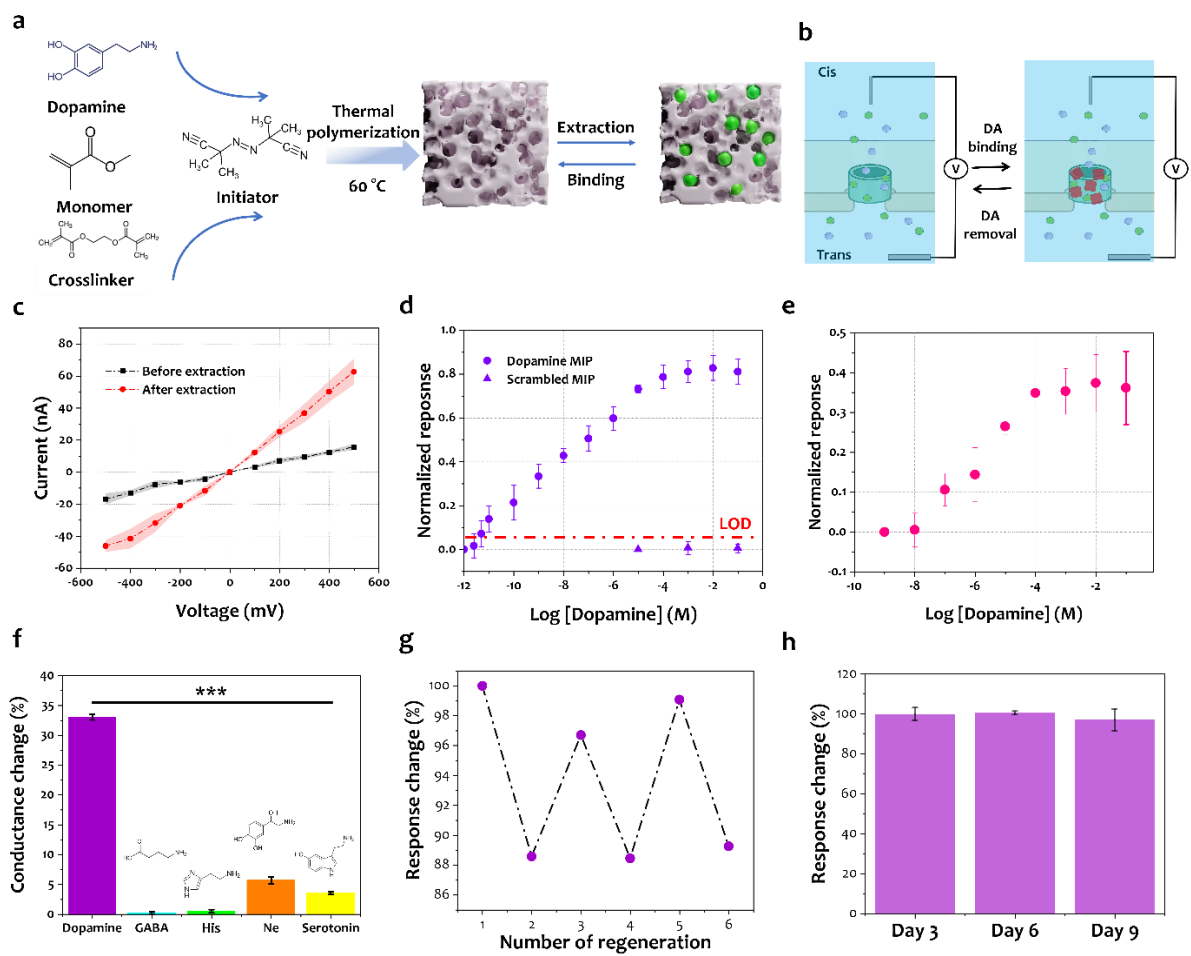

**Figure 2. Dopamine Sensor Characterization. (a)** Synthesis process of the MIP for dopamine detection. **(b)** Working mechanism of the MIP nanopore. **(c)** IV curve of the nanopore before and after the template removal in 1× phosphate buffered silane (PBS), the increase in conductance indicates the successful removal of the dopamine template, leaving behind selective binding cavities for dopamine. for N = 3 independent sensors **(d)** Calibration curve for dopamine detection in the range of 0.1 pM to 100 mM in 1x PBS. Each point is an average of N = 3 independent sensors. (N=3). The limit of detection (signal at zero analyte concentration plus 3 times its standard deviation) is shown by the red dotted line. **(e)** Calibration curve for dopamine detection in the range of 0.1 pM to 100 mM in neurobasal medium (N=3). **(f)** Selectivity test of the dopamine MIP nanopore against 1 nM of different analytes, dopamine, GABA, histamine, norepinephrine (NE), and serotonin in neurobasal medium [one-way ANOVA, ***$p < 0.0001$]. **(g)** The recovery of the MIP nanopore in presence of 1 uM dopamine. **(h)** Stability test of the dopamine MIP nanopore for up to 9 days (N=3).

## *Multiplexed Detection of Neurotransmitters*

Multiplexed sensing of biomolecules in real time offers significant advantages over individual target sensing. [34,58] For instance, in complex biological environments, where multiple neurotransmitters interact and regulate neural activity, monitoring their levels collectively provides a more comprehensive understanding of neural function. [13–15] After validating the performance of each MIP nanopore individually, the DA-MIP, GABA-MIP, and His-MIP were integrated onto a single $Si_3N_4$ membrane for multiplexed detection, as shown in Figure 3a. A significant challenge in developing a nanopore-based multiplexed platform lies in preventing cross-talk between the nanopores. In conventional configurations, the three nanopores share the same *cis* and *trans* chambers, rendering distinguishing signal from the individual nanopores difficult due to interference when multiple neurotransmitters are present in the solution (Fig. S19 - SI). To overcome this challenge, a new approach was introduced using conductive hydrogels. These soft, water-rich polymer networks not only support ionic conductivity but also allow precise, site-specific deposition through micro-dispensing techniques, as previously demonstrated. [59–61] By depositing micro-droplets of conductive hydrogel onto each nanopore (Figure 3a and 3b), the hydrogel served as individual micro-transducing chambers, providing localized ion transport channels. To reduce the inter-hydrogel distance and thereby increase the spatial resolution of the sensing platform, the deposition humidity was optimized. This optimization was critical as humidity directly influences droplet size, which in turn controls the spacing between the nanopores. [61,62] Figure S20 (SI) illustrates the diameter of droplets dispensed at different humidities; specifically, a chamber humidity of 38.35% RH consistently yielded 2.4 µm diameter droplets on the $Si_3N_4$ membrane, though sub-1 µm droplets are also achievable (Fig. S20 - SI). Hence, the use of micro-chambers enabled the conductance from each nanopore to be measured independently, preventing cross-talk between the signals originating from adjacent nanopores.

The incorporation of conductive hydrogels into the system thus allowed for reliable multiplexed detection with high spatial resolution of few µm by isolating the ionic signals from each nanopore. The bulk conductivity of a hydrogel film was measured to be ≈ 1.6 S/m, confirming the hydrogel's ability to facilitate ion transport, a critical factor for isolating signals. Furthermore, the ionic conductance of hydrogel-coated nanopore was measured by recording IV curves in 1× PBS, the linearity and slope of the IV curves provided insight into the hydrogel's inherent conductivity and possible changes in conductivity resulting from analyte interaction (Figures 3c and S21).

The performance of the multiplexing platform was then investigated by introducing three neurotransmitters, dopamine, GABA, and histamine, at different times into the shared trans chamber while measuring the current from each nanopore separately via the hydrogel-based micro-cis chambers. Notably, the measurement was performed in neurobasal medium to demonstrate the ability to multiplex in a complex environment containing amino acids and proteins. As illustrated in Figure 3d, each nanopore exhibited a selective response to its respective neurotransmitter. The DA-MIP nanopore showed a ~25% reduction in ionic current when exposed to dopamine, while the GABA-MIP and His-MIP nanopores exhibited no significant changes under the same conditions. Similarly, the GABA-MIP nanopore responded specifically to GABA, and the His-MIP nanopore demonstrated a significant current change only in the presence of histamine. Notably, the dopamine response in neurobasal was slightly lower than in PBS (~30%, Figure 2), which we attribute to the more competitive environment and partial nonspecific adsorption of proteins and amino acids in neurobasal medium [20,63]. Moreover, the sensing speed was measured to be around 100 ms (Fig. S22 - SI).

## *Selective Release of Neurotransmitters*

Another application of the MIP-nanopore is the selective release of neurotransmitters for targeted drug delivery systems [48,64–66]. Precise control over the release of neurotransmitters and small molecules can enable effective modulation of synaptic activity, addressing imbalances that contribute to conditions like Parkinson's disease, epilepsy, depression, and cancer treatment [57,67–72]. However, the reported delivery systems lack precise control on the release, selectivity, as well as the spatial resolution.

The ability to selectively release bound neurotransmitters from their respective MIP cavities is achieved through controlled electrical pulses. The release process is based on the disruption of non-covalent interactions between the MIP and the neurotransmitter, causing the neurotransmitter to be ejected from the binding cavity and released into the surrounding medium [35,64,66,73]. In Fig. 3e, the setup for selective release is shown, highlighting how distinct nanopores are individually triggered to release their bound neurotransmitter upon electrical stimulation. To investigate the ability of the platform to be used for controlled release, the release profile of the two different targets was investigated in 1x PBS. The release was triggered by 1 V for 30 s on each MIP to disrupt the binding of the neurotransmitters and then UV-vis spectroscopy was employed to measure the concentration in the buffer using 280 nm for dopamine for nanopore one and 480 nm for doxorubicin (DOX) for nanopore 2 (the DOX was chosen for having a shifted absorption peak). After 5 pulses approximately 80% of the loaded dopamine was released in the buffer (Figure. 3f).

As shown in Figure 3g, the nanopore's conductance initially was approximately 11 nS. Upon applying a reverse voltage, the conductance gradually increased to around 14 nS, indicating the release of dopamine molecules from the nanopore. Reapplying the original voltage resulted in a decrease in conductance, consistent with re-trapping of dopamine within the nanopore. This sensing and release cycle was repeated six times, demonstrating the platform's capability for reversible, closed-loop operation. The ability of multiplex detection and release capabilities provides a foundation for closed-loop systems, where real-time sensing of targets levels triggers the release of specific molecules in response to changing physiological conditions. This integration of detection and biomolecule release paves the way for smart, responsive systems in neurochemical regulation, significantly advancing the field of sensing and therapeutics. Further analyses of the underlying release mechanism are provided in the Supporting Information (Note 7-9).

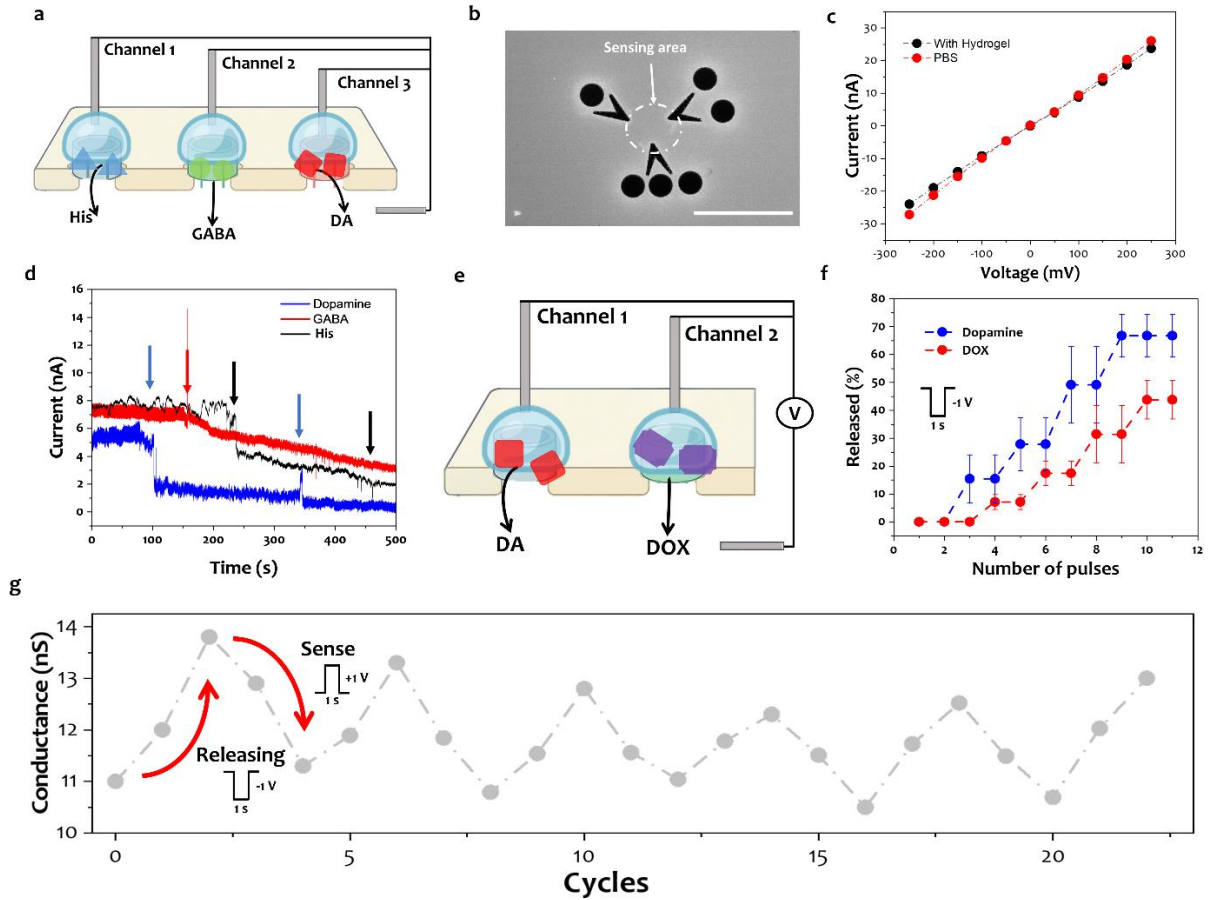

**Figure 3. Multiplex Sensing and Releasing of Dopamine, GABA, and Histamine. (a)** Illustration of the multiplexed sensing platform on a silicon nitride membrane. Each nanopore is functionalized with a specific MIP for dopamine, GABA, and histamine, then separate top electrodes consisting of conductive hydrogels allow simultaneous detection without cross-talk between the nanopores. **(b)** SEM image of the three nanopores on the same membrane, with (arrow + circle) markers to locate the nanopores for the hydrogel deposition. Scale bar 10 µm. (c) IV curve of the nanopore measure in PBS and with conductive hydrogel. (d) Ionic current changes for each neurotransmitter (dopamine, GABA, histamine) in a multiplexed system, showing distinct conductance reduction for each neurotransmitter upon binding in neurobasal medium, the arrows indicates the injection time of different neurotransmitters. (e) Illustration of the multiplexed delivery platform on a silicon nitride membrane. Each nanopore is functionalized with a specific MIP for dopamine and doxorubicin (DOX). (f) Release profile of the loaded molecules from the MIP-functionalized nanopores via electrical pulses in 1x PBS (N=3). (g) Conductance change over time in the closed-loop platform for dopamine sensing and release, demonstrating the system's ability to detect and subsequently release the trapped (sensed) dopamine molecule back into the analyte volume.

## Neurotransmitters for bio-computation

The third block to perform in-edge computing is the implementation of logic gates such as NAND, NOR, and NOT using two neurotransmitters, such for example dopamine and GABA [35,74–77]. In this neurotransmitter-based logic platform, the presence or absence of the neurotransmitters determines a "0" or "1" input, and ionic conductance serves as the binary output. Specifically, a low conductance state corresponds to a logic "0", while a high conductance state represents a logic "1". These two states are achieved by the binding or absence of the neurotransmitters within their respective nanopores. When neurotransmitters bind to the MIP-functionalized nanopores, conductance is reduced, allowing for binary logic operations based on these neurotransmitter interactions (Figure. 4a) [78]. In the case of the NOT gate, a single neurotransmitter is required as input. If the neurotransmitter is absent (representing a "0"), the output is high conductance ("1"), and *vice versa*, thereby inverting the input signal. This basic logic function showcases the system's capacity to process unitary operations at the molecular level (Figures. 4b). For more complex operations, the platform supports NOR and NAND gates, which are essential for constructing digital circuits. In these gates, two inputs, dopamine and GABA, are required, utilizing the DA-MIP and GABA-MIP functionalized nanopores on the same membrane. In the NAND gate, the output is "0" (low conductance) only when both inputs, dopamine and GABA, are present and bound to their respective MIP-functionalized nanopores. If either neurotransmitter is absent or unbound, resulting in higher conductance, the output is "1" (Figures. 4c).

Conversely, the NOR gate produces an output of "1" (high conductance) only if both inputs — dopamine and GABA — are absent. If either or both neurotransmitters are present, the output becomes "0". The previously developed DA-MIP and GABA-MIP systems demonstrate high selectivity, such that the presence of only one neurotransmitter does not block both MIP-functionalized nanopores simultaneously. This selectivity allows for thresholding, where a conductance value below a set threshold is considered "off" and above it is "on" [79]. However, the use of thresholds can result in current accumulation when multiple logic gates are connected in series, potentially leading to computational errors. To address this issue and create a more reliable logic system, we designed a new MIP-functionalized nanopore sensitive to both dopamine and GABA. In this configuration, the presence of either neurotransmitter results in full blockage of both nanopores, ensuring an output of "0" whenever either or both neurotransmitters are present (as shown in Fig. 4d). This approach avoids the issue of accumulated current and allows for more robust computational operations, further advancing the potential of neurotransmitter-based logic systems in neuromorphic and bio-inspired computing. These neurotransmitter-driven logic operations represent a novel method for integrating biological signals into computational frameworks. The use of neurotransmitters to modulate ionic conductance introduces a biologically relevant method to perform logic operations, which can be applied in bio-computing systems, neural interfaces, and real-time bio-signal processing.

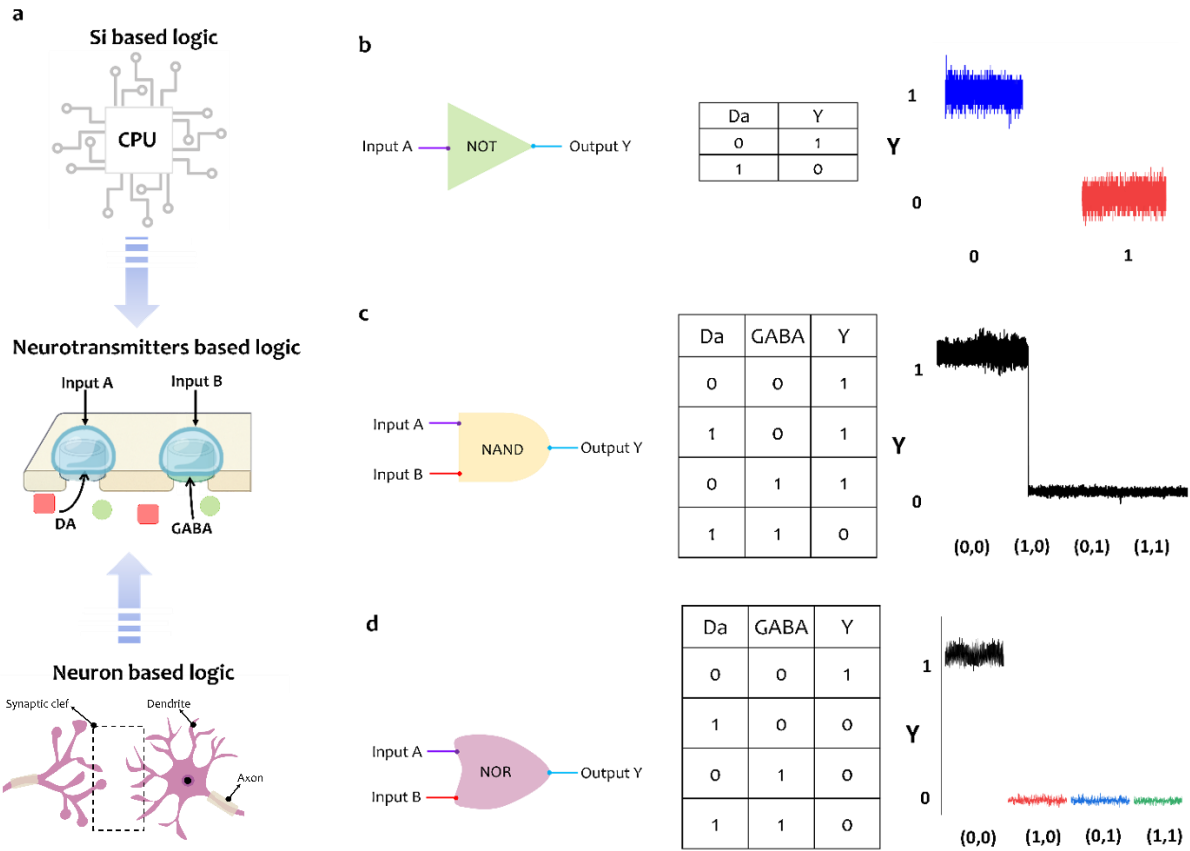

**Figure 4. MIP-Nanopore System for Logic Gate Operations Using Dopamine and GABA:** **(a)** Illustration of the working mechanism of the MIP-nanopore logic gate, functionalized with MIP specific to dopamine and GABA, the two neurotransmitters act as input A and B. **(b-d)** Depict the implementation of logic gates using the MIP-nanopore system. **(b)** NOT gate is achieved by the binding of dopamine, which blocks the nanopore and results in an output of "0" (low current), whereas the absence of dopamine yields an output of "1" (high current). **(c)** NOR gate, the presence of either dopamine or GABA results in an output of "0," while the absence of both neurotransmitters generates an output of "1." **(d)** The NAND gate produces an output of "1" if at least one neurotransmitter is absent, maintaining a high ionic current.

## Performing in-edge computing

Figure 5a and Fig. S26-S28 (SI) illustrate a dual-chamber system integrating sensing, in-edge computing, and controlled release mechanisms for the selective detection and regulation of neurotransmitters, here dopamine and GABA. The setup consists of a sensing volume for detecting neurotransmitters (acting as the A and B inputs) and a releasing volume for the controlled release of dopamine, connected *via* a MIP-nanopore logic gates. In-memory sensing enables local detection of dopamine and GABA, while edge computing processes the sensory data to determine the release response (releasing or stop releasing dopamine). Fig. 5c shows the release pattern in case of a NOT gate, in absence of dopamine in the sensing volume led to an increase in dopamine concentrations in the released volume, however once dopamine was injected in the sensing volume (reaching the threshold) the release of dopamine was stopped, which eventually saturates. In the case of NOT gate, the system's operational logic is summarized in the Table in Fig. 5d, dopamine release occurs only when dopamine and GABA are absent, while the presence of GABA, alone or with dopamine, halts the release. The third case is the NAND gate, where the release of dopamine halts only in the presence of both dopamine and GABA. However, the presence of only one of the two targets only decreases the release but does not fully stop it.

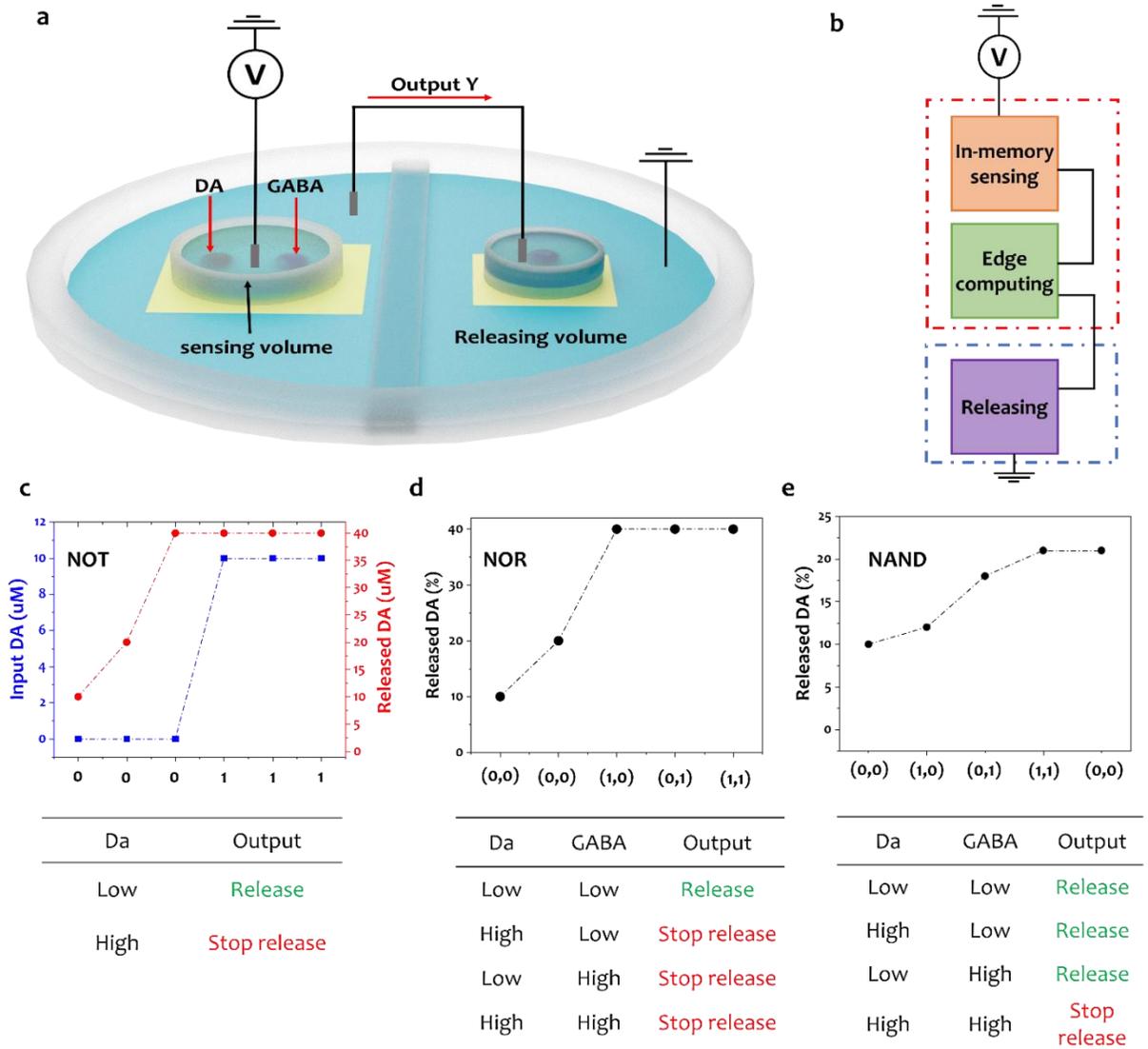

**Figure 5. MIP-Nanopore Platform for In-Edge Computing and In-Memory Sensing. a)** Illustration of the setup used for in-edge computing and releasing by employing MIP-nanopores. b) block diagram of different parts of the ionic in-memory sensing. (c–e) In-edge computing logic operations demonstrated with the MIP-nanopore platform. In each panel, the blue curve represents the input DA concentration in the sensing volume, and the red curve represents the DA released into the releasing volume. The release behavior corresponds to the specific logic gate implemented: (c) NOT gate: release occurs only when DA is absent (input = 0), and stops when DA is present (input = 100 uM = 1). (d) NOR gate: release occurs only when both DA and GABA are absent; the presence of either neurotransmitter stops release. (e) NAND gate: release stops only when both DA and GABA are present simultaneously; if one or both are absent, release continues.

# Conclusion

In this study, we demonstrated a radically novel platform that combines solid-state nanopores with MIPs for the selective multiplexed detection and controlled release of neurotransmitters namely, dopamine, GABA, and histamine. The individual MIP-nanopores were combined into a single platform, enabling, for the first time, multiplexed detection of different biomolecules. This integrated platform was achieved by functionalizing distinct nanopores with neurotransmitter-specific MIPs with separate electrical contacts using a conductive hydrogel to prevent signal cross-talk between the different nanopores. This system could differentiate neurotransmitters in real time, an essential feature for investigating complex neurochemical interactions in the brain. Moreover, the platform's capacity to selectively release neurotransmitters *via* controlled electrical pulses holds great promise for therapeutic applications. Additionally, we demonstrated that this system can be used to construct logic gates (NOT, NOR, NAND) based on the presence of specific neurotransmitters. Subsequently, the sensing, releasing, and logic operations were integrated to form an in-edge computing platform. This platform is capable of sensing neurotransmitters, performing logic-based computation, and triggering the controlled release of molecules in response. In conclusion, the MIP-nanopore platform we developed offers a versatile tool for multiplexed neurotransmitter sensing, selective release, and bio-computation, with vast potential for applications in neurochemical monitoring, diagnostics, neuromorphic systems, and therapeutic interventions.

# Materials and Methods

## 2.1 Materials

Sigma-Aldrich was the main supplier for the chemicals used in this work unless otherwise noted. All measurements utilized phosphate buffer saline (PBS) at 1× concentration (137 mM NaCl, 2.7 mM KCl, 10 mM $Na_2HPO_4$, 1.8 mM $KH_2PO_4$) and pH 7.4 (ThermoFisher Scientific AG) as received.

### Fabrication of Solid-State Nanopores

Freestanding $Si_3N_4$ membrane chips were prepared following a standard membrane fabrication procedure. In particular, an array of square membranes was prepared on a commercial double-sided 100 nm LPCVD $Si_3N_4$ coated 500 μm Si wafer via UV photolithography, following reactive ion etching and subsequent KOH wet etching. Afterward, nanopores featuring a diameter of 40 ± 10 nm were fabricated using a focused $Ga^+$ ion beam. Precise adjustments to milling times were made to achieve the optimized pore diameter.

### Surface Functionalization

After FIB drilling, the chips were thoroughly cleaned using ethanol/IPA followed by rinsing with deionized water and drying under a nitrogen stream and finally, oxygen plasma for 5min. Afterwards, the cleaned nanopores were surface modified by thermal deposition of 10 nm $SiO_2$, followed by APTES coating with atomic layer deposition.

### Synthesis of MIPs

The MIPs were synthesized by *in situ* polymerizations of the target template. For instance, dopamine-MIP, methacrylic acid monomer (MAA), acrylamide (ACM), and *N,N′*-methylenebis(acrylamide) (MBAA) at a 1:2:2:10 molar ratio. First, the monomeric components, MAA and ACM (both 0.3 M), and the cross-linking monomer, MBAA (1.5M) were dissolved in a methanol/water mixture (25 mL, 4:1, v/v) along with the initiator, acrylamide (ACM), 2,2′-azobis(1-methyl-propionitrile) (AIBN) (0.284 gr), and dopamine (0.15M). Subsequently, the mixture was spin coated on the nanopore and copolymerized at 60 °C for on a hot plate. Once the polymerization was complete, the neurotransmitter template was removed from the polymer matrix by extensive washing with solvents (methanol/ACN), leaving behind selective binding cavities specific to the neurotransmitter. At the end, three different MIPs were synthesized, Dopamine-MIP (DA-MIP), GABA-MIP, and Histamine-MIP (Hist-MIP). Moreover, control MIPs were also synthesized without adding the target neurotransmitters. In the case of multiplexing devices an inkjet molecular printer " BioForce NanoEnabler system" was used to deposit MIP selectively on the nanopores and then polymerized, afterwards the conductive hydrogel were deposited. Finally, the prepared samples were immersed in water until used to improve wettability and to dissipate bubbles.

*Characterization of MIPs and MIP-nanopores*

AFM data was taken by non-contact mode using an AFM system XE-100, Park Systems.

**Quartz Crystal Microbalance**

KSV QCM-Z500 microbalance was used for QCM-D measurements, and dopamine-MIP was coated on gold chips. Before coating, the chips underwent a rigorous cleaning procedure consisting of 2 min sonication in 2-propanol, acetone, and Milli-Q water. Later on, the chips were dried using nitrogen and then oxygen plasma was cleaned for 10 min and finally, the dopamine-MIP was spin-coated. The 7th harmonic was used for the analysis of the dissipation, frequency representation.

*Conductance Measurements for Neurotransmitter Detection*

The detection mechanism of the nanopore platform is based on monitoring changes in ionic current. Conductance measurements were performed in 1x PBS buffer solution by applying a transmembrane across the SiN membrane, and the resulting ionic current was measured using the same instrument/reader from Elements srl. Changes in conductance were used to detect neurotransmitter binding to the MIP cavities. A decrease in conductance indicated neurotransmitter binding, while an increase in current after washing with solvent indicated the successful removal of the neurotransmitter. the LOD and LOQ were calculated using the following formula:

$$LOD = \frac{3.3\sigma}{S}$$

$$LOQ = \frac{10\sigma}{S}$$

*Where:* $\sigma$ = the standard deviation of the response

$S$ = the slope of the calibration curve

*Multiplexed Detection Setup*

For multiplexing, the three different MIPs (DA-MIP, GABA-MIP, and histamine-MIP) were immobilized on spatially separated regions of a single SiN membrane. Hydrogels were applied to the surface of each nanopore region to enhance ionic conductivity and prevent cross-talk between neighboring nanopores. The hydrogels served as conductive media that localized the ionic current changes to specific nanopores, enabling independent detection of each neurotransmitter. First, A solution of polyvinyl alcohol (PVA, 8 wt%) was first prepared by dissolving PVA in 1 x PBS at 95 °C. To this solution, sodium alginate (SA, 3 wt%) was introduced and the mixture was stirred at 85 °C for 3 h, yielding a homogeneous PVA–SA blend. Separately, an aqueous solution of sodium tetraborate (STB, 4 wt%) was prepared under stirring at 85 °C for 1 h. The STB solution was then combined with the PVA–SA mixture, followed by continued stirring at 85 °C, resulting in the formation of a conductive hydrogel. Afterwards, BioForce Nano Enable system was used to first deposit MIP for each neurotransmitter target on the specific location using the markers (Figure. 2b) followed by diposition of hydrogels on top of each MIP-nanoproe. The platform was tested by exposing nanopores to a solution containing a mixture of dopamine, GABA, and histamine. The ionic current response from each nanopore was measured independently by using the two channels of keithley 2600b plus the Elements reader.

*Selective Release of Neurotransmitters*

In addition to detection, the nanopore platform was designed to enable the selective release of neurotransmitters from the MIP cavities. This was achieved by applying controlled electrical pulses to disrupt the non-covalent interactions between the neurotransmitter and the MIP binding cavities. Electrical pulses were delivered to the nanopores using a voltage-controlled setup previously used for multiplex sensing. The release profiles were tracked by measuring the absorption change of the electrolyte using a UV–visible spectrophotometer (CARY200 Scan, Varian), over multiple pulses. The data are expressed as a cumulative percentage of triplicate experiments.

The loading capacity and entrapment efficiency of the targets were calculated using the following expressions[80]:

$$\text{drug loading capacity \%} = \frac{\text{amount of target in sensing solution}}{\text{total amount of DOX for loading}} \times 100$$

**Fabrications of Nanoheater-Embedded Nanopores**

A photoresist layer (AZ5214E, MicroChemicals) was first spin-coated onto SiN membranes (100 nm thick, 5 × 5 mm chips). Alignment markers consisting of Au/Ti (30/5 nm) were then defined at the edges of the membrane by standard photolithography. These markers served as references for positioning the nanoheater. A spiral heater pattern was subsequently written by Dual Beam FIB-SEM system using PMMA A2 (MicroChem) as the resist. After development, a 20 nm Pt film was deposited by sputtering, followed by lift-off to yield the Pt nanowire structure functioning as the local heater. A nanopore was then drilled at the center of the Pt spiral using FIB milling. To minimize parasitic leakage currents during measurements in electrolyte, the chip surface was conformally coated with a 20 nm $SiO_2$ layer by atomic layer deposition, leaving only the square pads exposed for electrical connection.

**Nanoheater Calibration**

A nanoheater-embedded nanopore chip was mounted on a microfluidic chip, then the current through the Pt nanowire under 0.05 V using a sourcemeter (keithley 2600b). Afterwards, the platinum resistance was measured under a voltage ramp in 1x PBS using a hotplate. Therefore, a calibration curve was constructed temperature vs nanoheater's resistance.

**SUPPORTING INFORMATION**

**Hybrid nanopores for real-time multiplexed sensing/delivery of biomolecules and logic computing**

*Ali Douaki,\* Shukun Weng, Silvia Dante, Nako Nakatsuka, Makusu Tsutsui, Roman Krahne, and Denis Garoli\**

A.Douaki, S. Weng, G. Lanzavecchia, A. Sapunova, S. Dante, R. Krahne, D. Garoli

Istituto Italiano di Tecnologia, Via Morego 30, Genova 16163, Italy

N. Nakatsuka

Laboratory of Chemical Nanotechnology, EPFL, Genève, 1202, Switzerland

M. Tsutsui

The Institute of Scientific and Industrial Research, Osaka University, Mihogaoka 8-1Ibaraki Osaka 567-0047, Japan

A.Douaki, D. Garoli

Dipartimento di scienze e metodi dell'ingegneria Università di Modena e Reggio Emilia Via Amendola 2, Reggio Emilia 42122, Italy

E-mail: ali.douaki@unimore.it; denis.garoli@unimore.it

**Note 1. Device characterization**

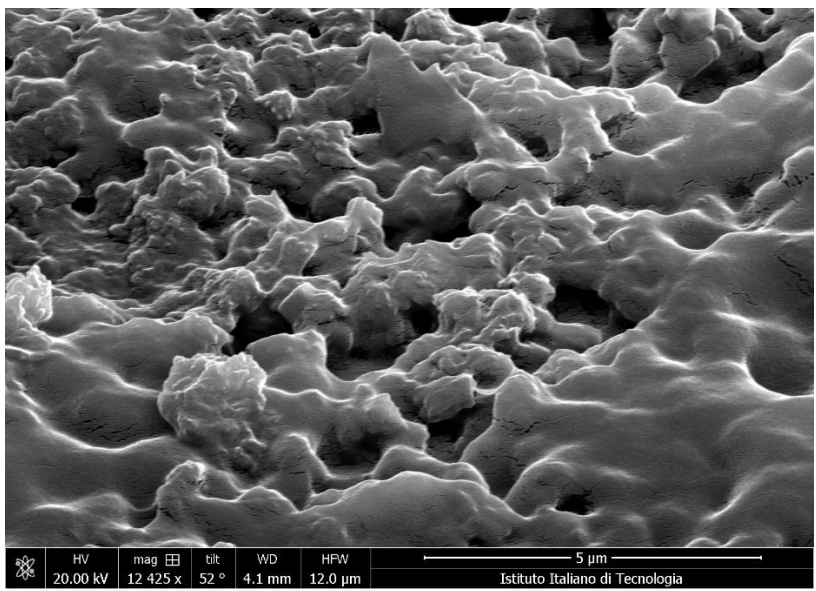

**Figure S1.** Scanning electron microscopy (SEM) image of the MIP for dopamine with the porous structure of the MIP, Scale bar, 5 um.

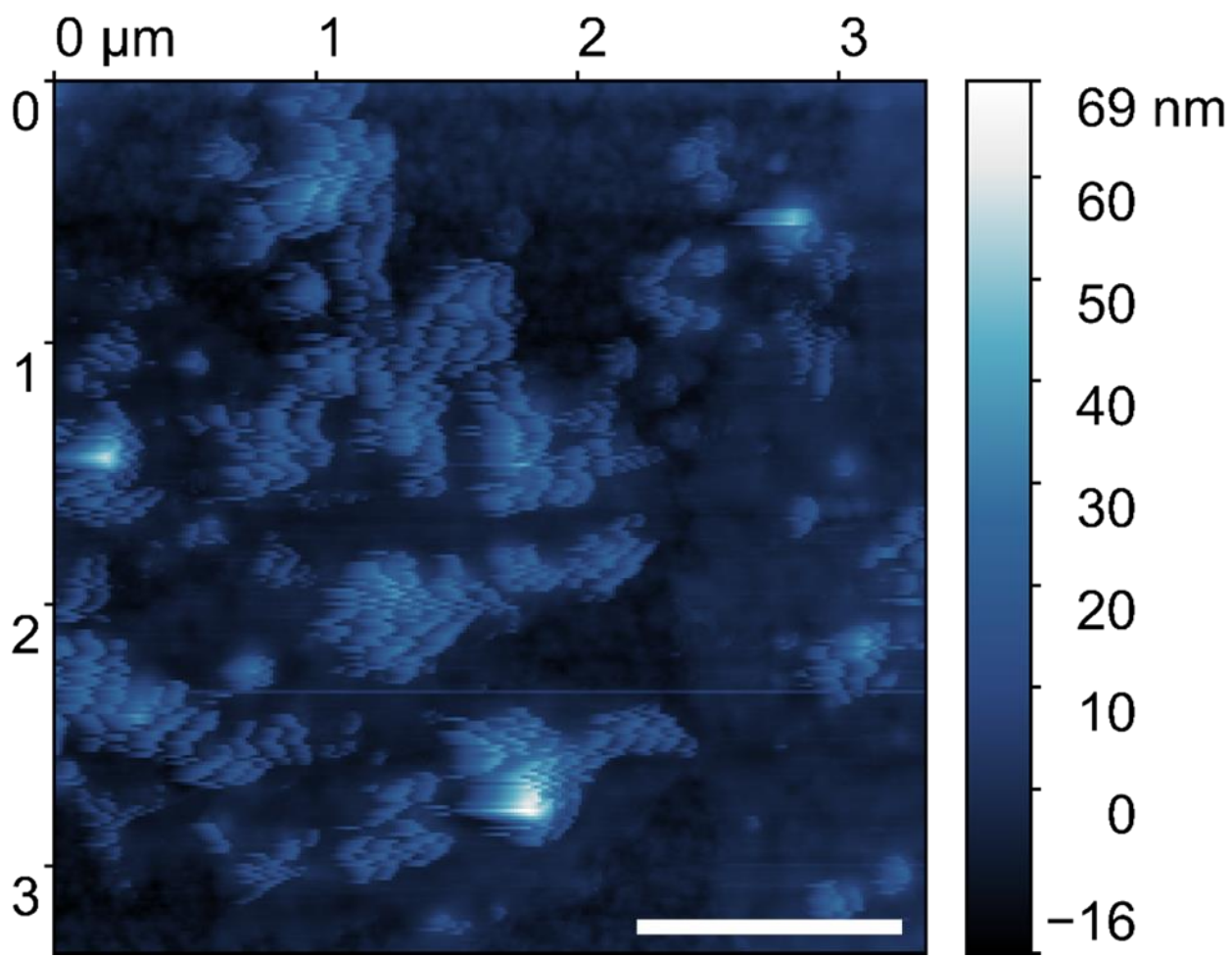

**Figure S2. Atomic force microscope** (AFM) image of the MIP for dopamine. Scale bar, 1μm.

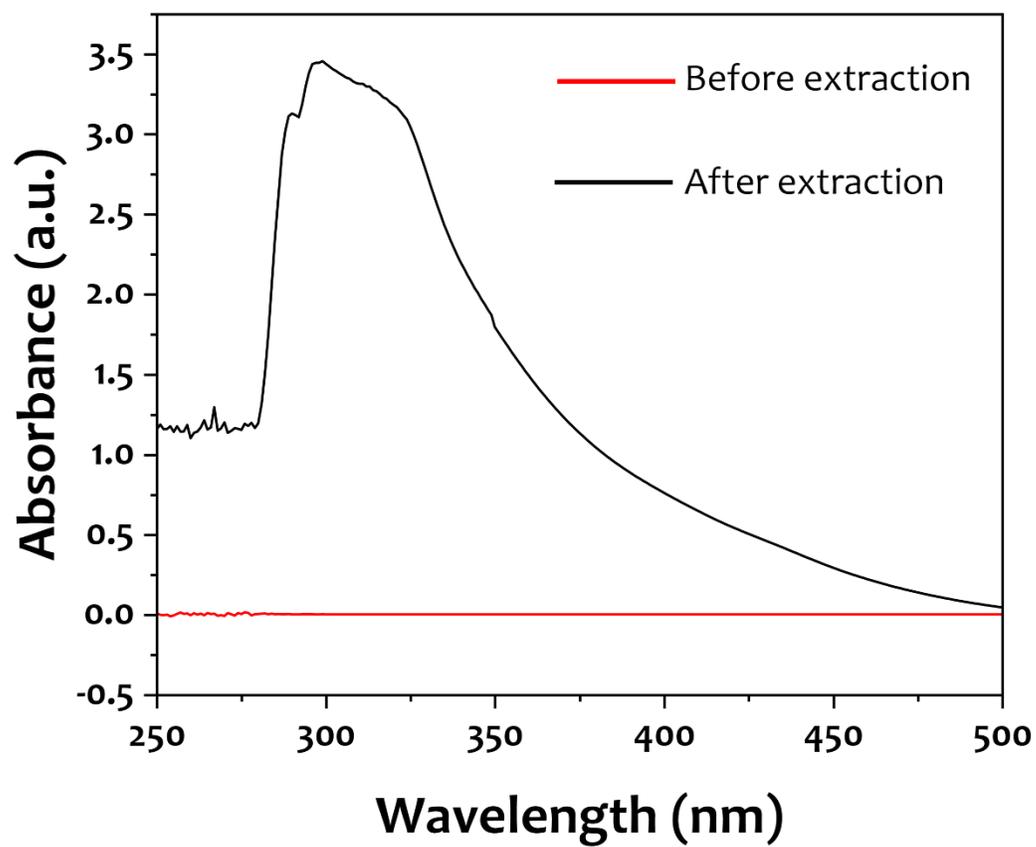

**Figure S3.** ultraviolet–visible spectroscopy (UV–vis) absorbance of dopamine MIP before and after template extraction

## Note 2. Measurement setup

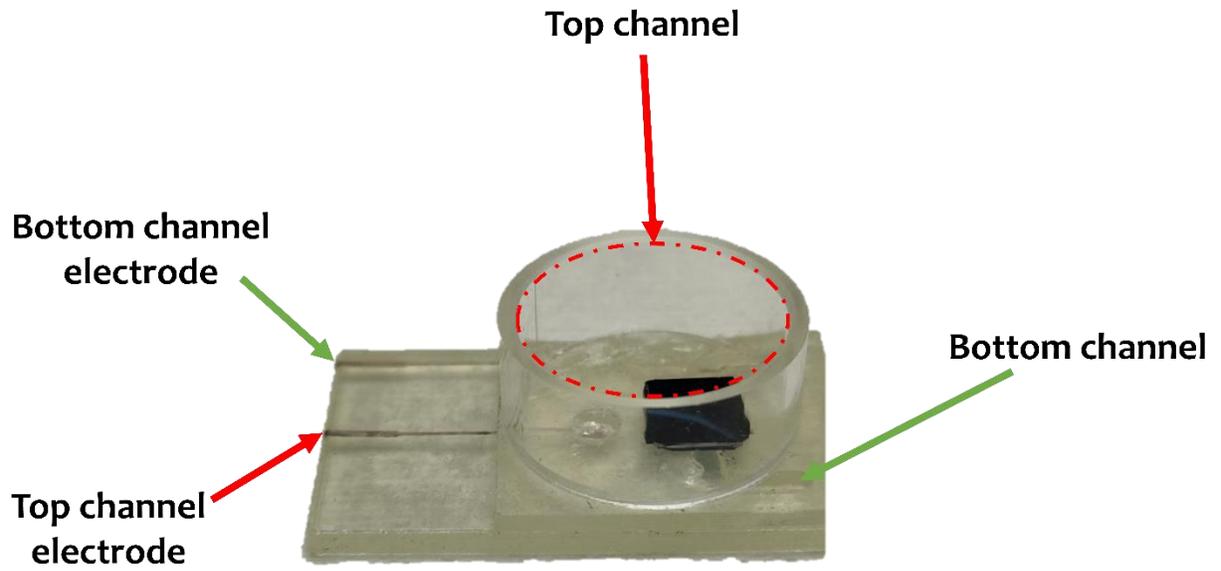

**Figure S4.** The microfluidic used in electrical measurements.

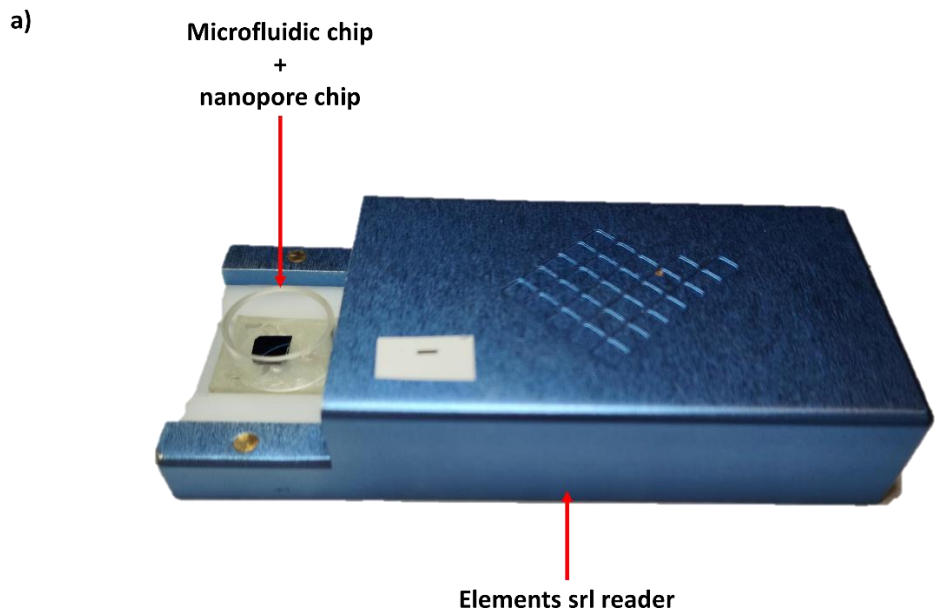

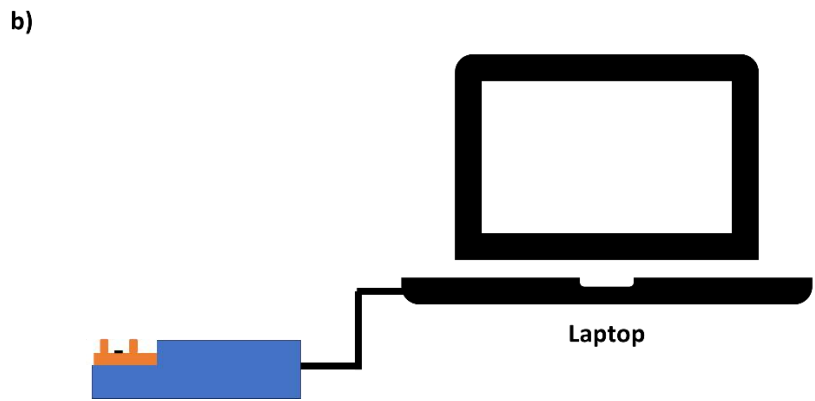

**Figure S5. (a-b)** The microfluidic chamber connected to electrical nanopore reader (Elements SRL).

**Note 3. Dopamine MIP-nanopore characterization**

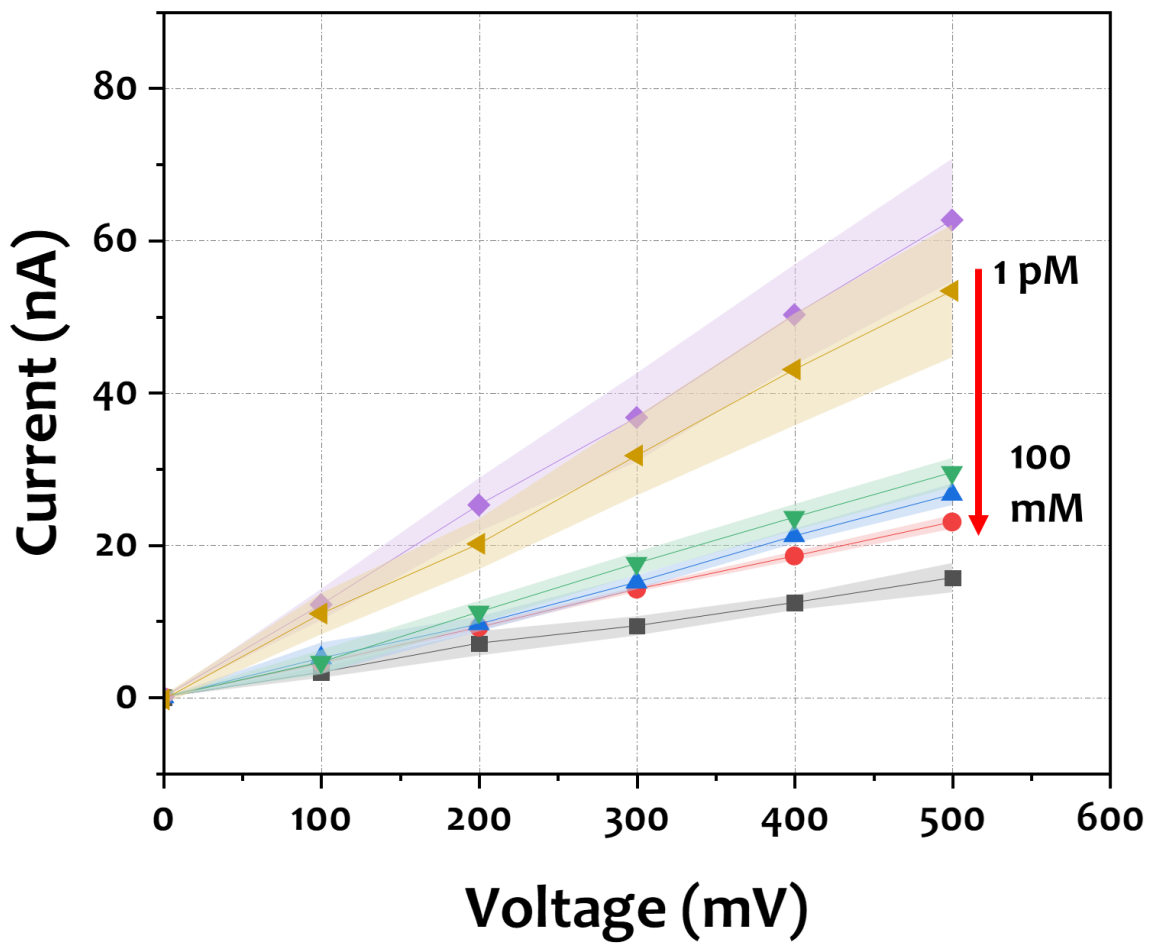

**Figure S6.** Concentration-specific response of MIP-dopamine sensors in 1× phosphate buffered silane (PBS) with increasing dopamine amounts observed in IV curves. Each point is an average of N = 3 independent sensors.

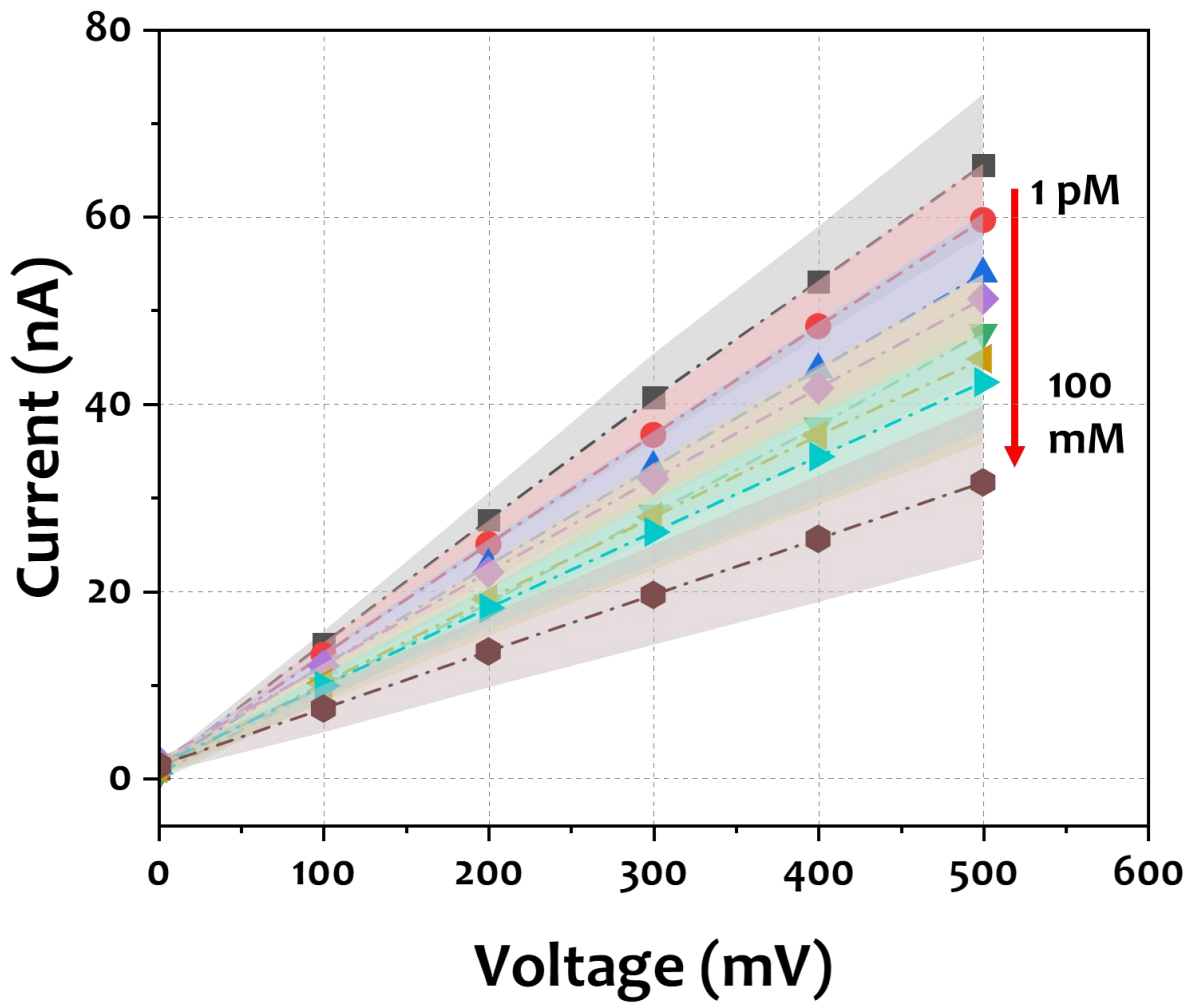

**Figure S7.** Concentration-specific response of MIP-dopamine sensors in neurobasal medium **with increasing dopamine amounts observed in IV curves. Each point is an average of N = 3 independent sensors.**

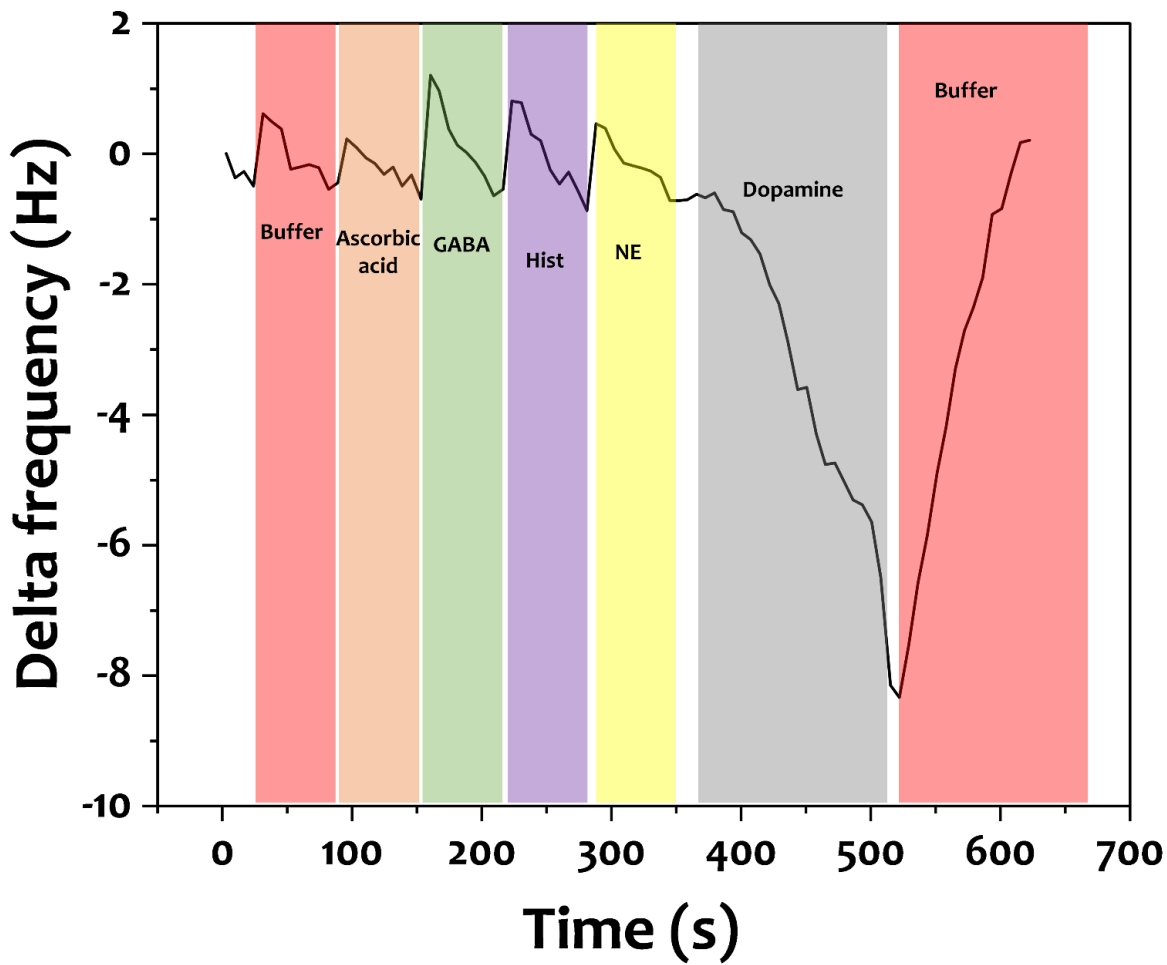

**Figure S8.** QCM-D selectivity analysis of the dopamine-MIP at the 5th overtone. Frequency shifts (Δf) were recorded upon sequential exposure to 1 nM of ascorbic acid, GABA, histamine, norepinephrine (NE), and dopamine in neurobasal medium prepared in 1× PBS, followed by buffer rinse at 500 s. A pronounced response was observed only for dopamine, confirming selective recognition

**Note 4. GABA MIP-nanopore characterization**

The sensitivity of the GABA MIP nanopore was assessed across a concentration range of 0.1 nM to 100 mM (Figure. s9). The sensor demonstrated a robust linear response, with an $R^2$ value of 0.96, indicating strong reproducibility and sensitivity within this range (Figure. S10). Additionally, GABA MIP nanopore achieved a LOD of 1.99 pM, making it well-suited for detecting GABA at concentrations relevant to neurochemical signaling and synaptic activity [1,2]. The applicability of this sensor depends on its deployment environment, as the extracellular level of GABA in the brain typically range from 200 nM to 1 µM under basal conditions and can spike to 10 to 100 µM during synaptic events. Therefore, careful calibration is essential for adapting the sensor to specific physiological or experimental contexts. Therefore, a sensor with a working range of 0.1 nM to 100 mM offers considerable flexibility for translational applications, particularly for monitoring GABA fluctuations within the brain, where concentrations vary between basal and synaptic states. Moreover, selectivity tests revealed less than 5% current reduction when exposed to dopamine, histamine, norepinephrine, and serotonin at 10 µM, confirming the sensor's high specificity for GABA (Figure. S11). Reusability tests showed that the sensor retained 92% of its initial sensitivity after 15 binding and release cycles, with consistent stability over a 9-day period (Figure. S12 and S13).

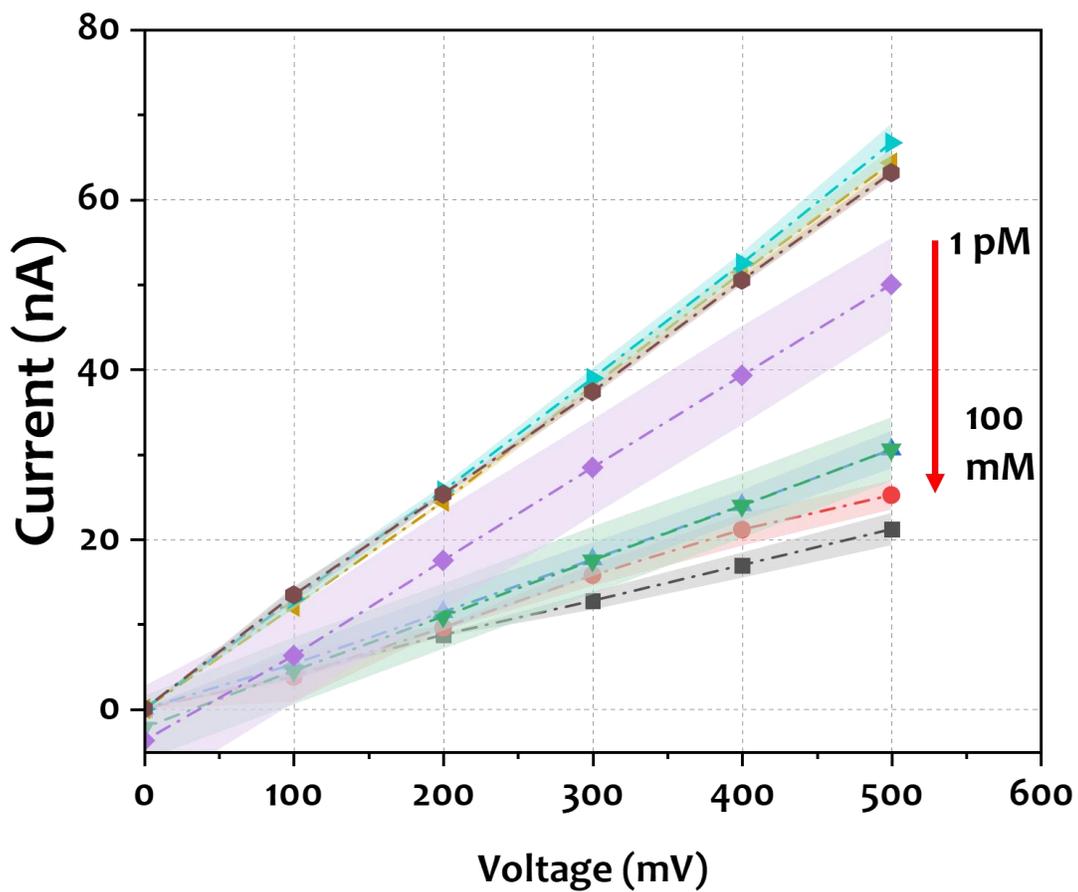

**Figure S9.** Concentration-specific response of MIP-GABA sensors in 1× phosphate buffered silane (PBS) with increasing GABA amounts observed in IV curves. Each point is an average of N = 3 independent sensors.

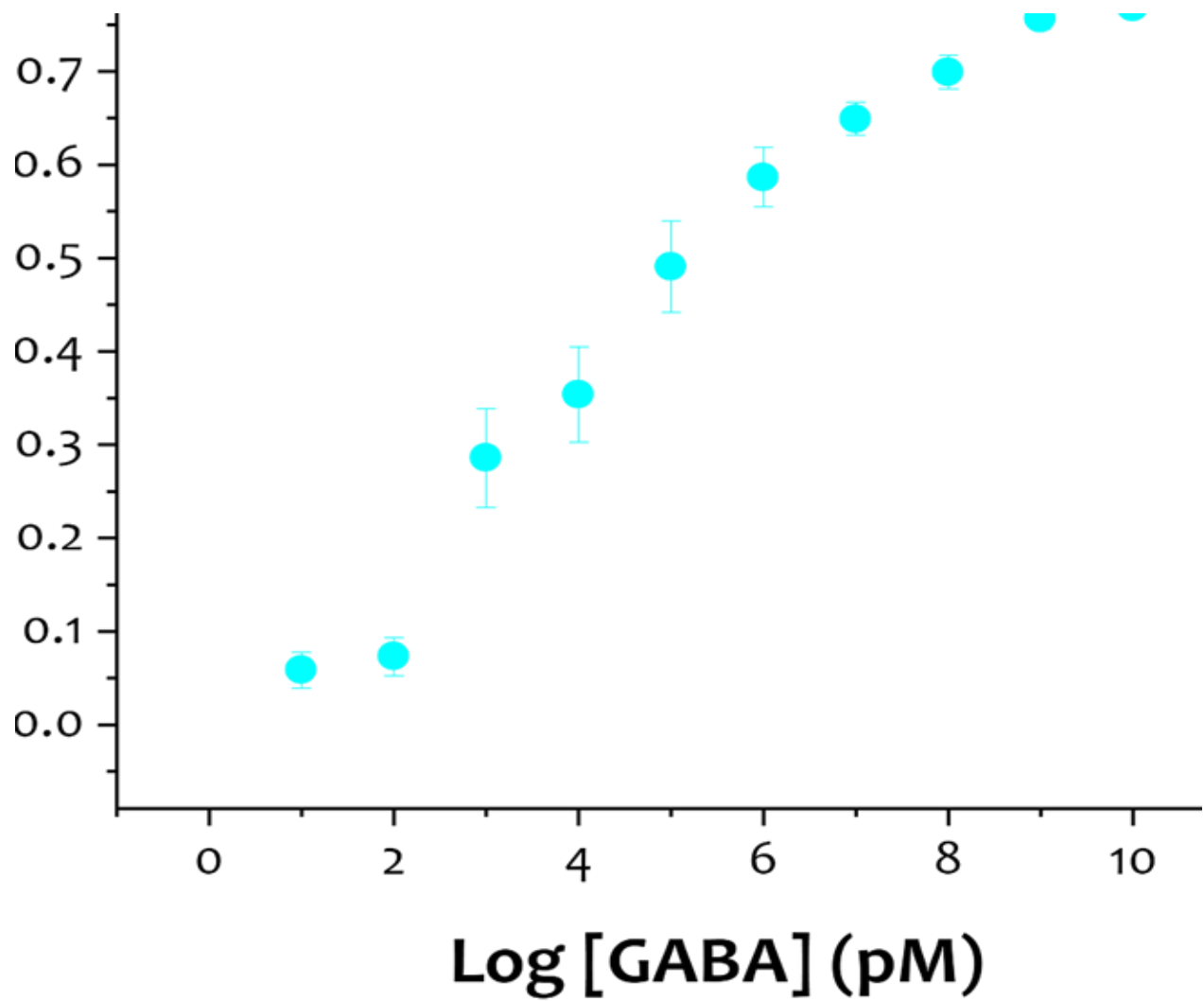

**Figure S10.** Calibration curve for GABA detection in the range of 0.1 pM to 100 mM in 1x PBS. Each point is an average of N = 3 independent sensors.

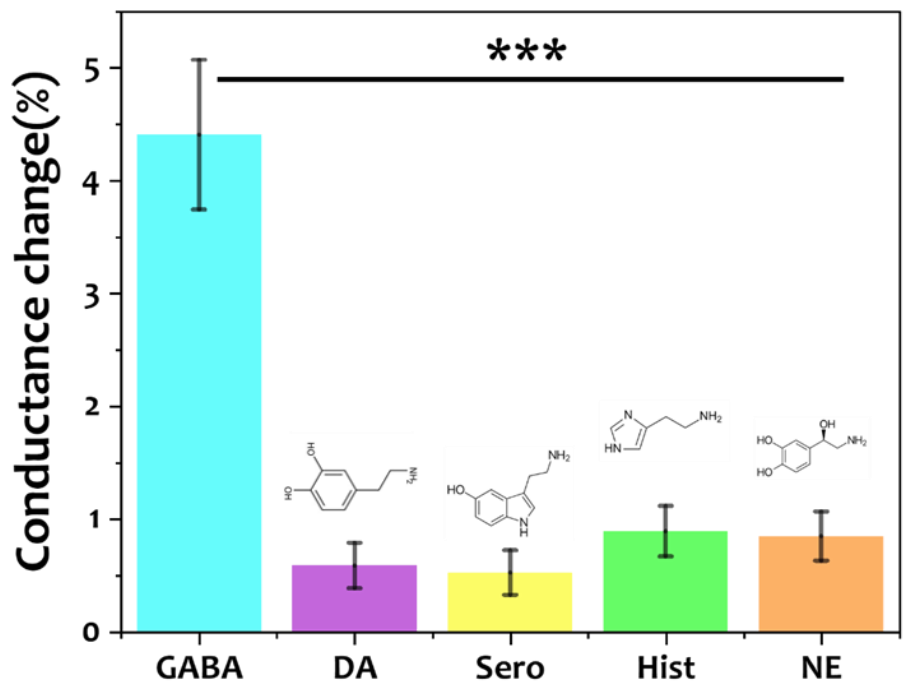

**Figure S11.** Selectivity test of the GABA-MIP nanopore against 1 uM of different analytes, dopamine (DA), gamma-aminobutyric acid (GABA), Histamine (Hist), norepinephrine (NE), and serotonin (Sero) in neurobasal medium. Each point is an average of $N = 3$ independent sensors. [one-way ANOVA, ***$p < 0.0001$]

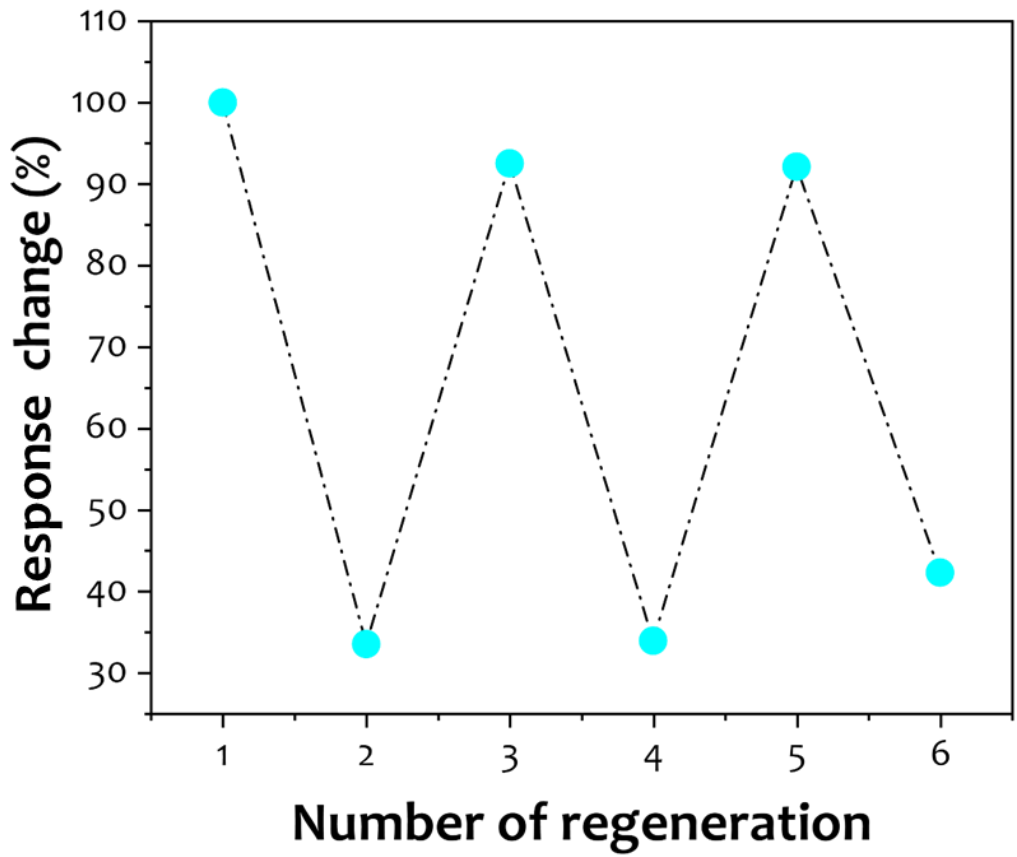

**Figure S12.** The recovery of the GABA-MIP nanopore in presence of 1 uM GABA in 1x PBS.

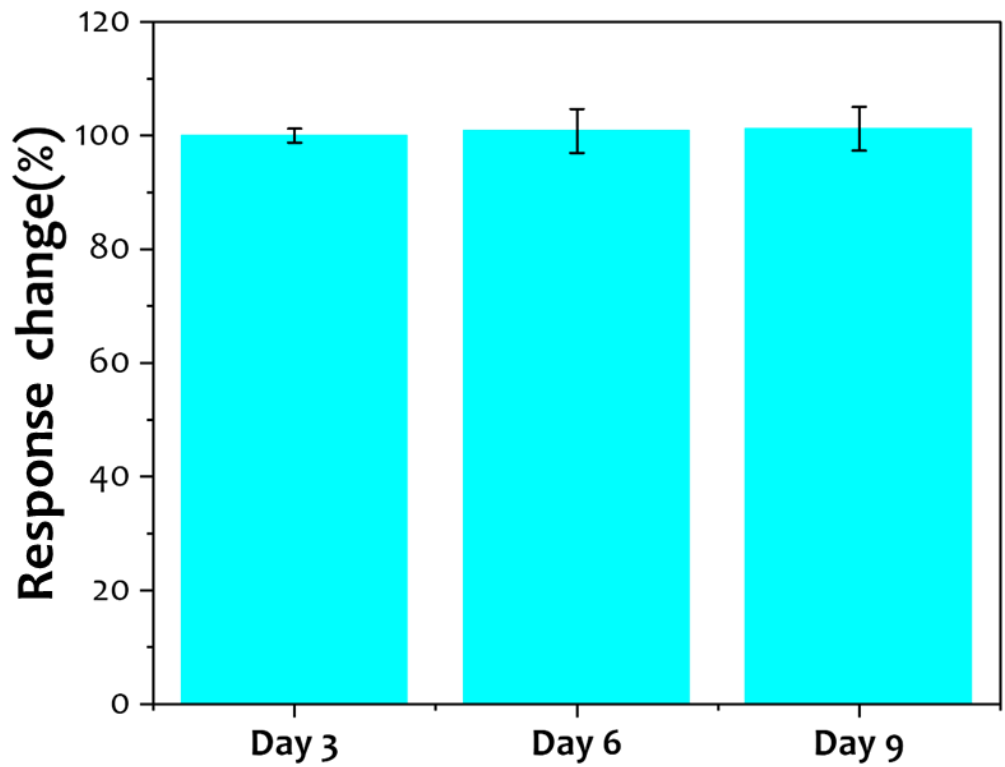

**Figure S13.** Stability test of the GABA MIP nanopore for days in 1x PBS. Each point is an average of $N = 3$ independent sensors.

## Note 5. Histamine MIP-nanopore characterization

The third MIP-nanopore sensor was developed for Hist detection. the sensor demonstrated a linear response to histamine concentrations ranging from 1 nM to 100 mM, with a coefficient of determination ($R^2$) of 0.97 (Fig. S14-S15). Moreover, the sensor showed a LOD of 2 nM, therefore, the sensor exhibits high sensitivity suitable for monitoring low concentrations of histamine relevant to immune responses or allergic reactions. Hist concentrations in the brain vary from low nanomolar to micromolar levels. Basal concentrations in human cerebrospinal fluid are typically around 1–20 nM, while localized levels can increase to several micromolar during heightened neuronal activity or in response to specific stimuli. Therefore, a sensor capable of detecting histamine within the low nanomolar to millimolar range is appropriate for monitoring both basal and physiologically elevated histamine levels in neurological studies. The histamine sensor exhibited minimal current changes (less than 5%) when exposed to 1 µM concentrations of dopamine, GABA, norepinephrine, or serotonin, further confirming its selectivity (Fig. S16). Long-term stability assessments revealed that the sensor retained 90% of its initial sensitivity after 15 cycles, with consistent performance over 30 days (Fig. S17-S18).

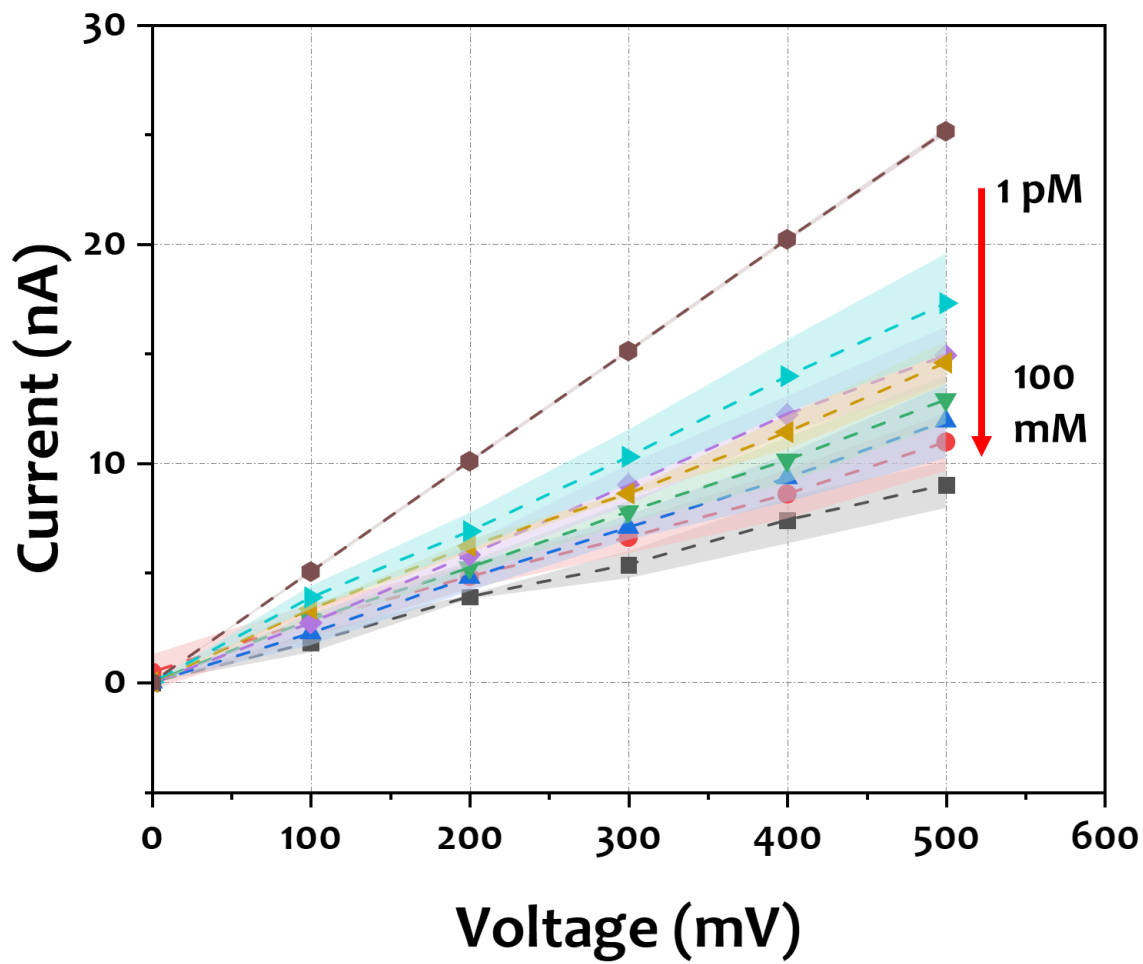

**Figure S14.** Concentration-specific response of MIP-histamine sensors in 1× phosphate buffered silane (PBS) with increasing histamine amounts observed in IV curves. Each point is an average of N = 3 independent sensors.

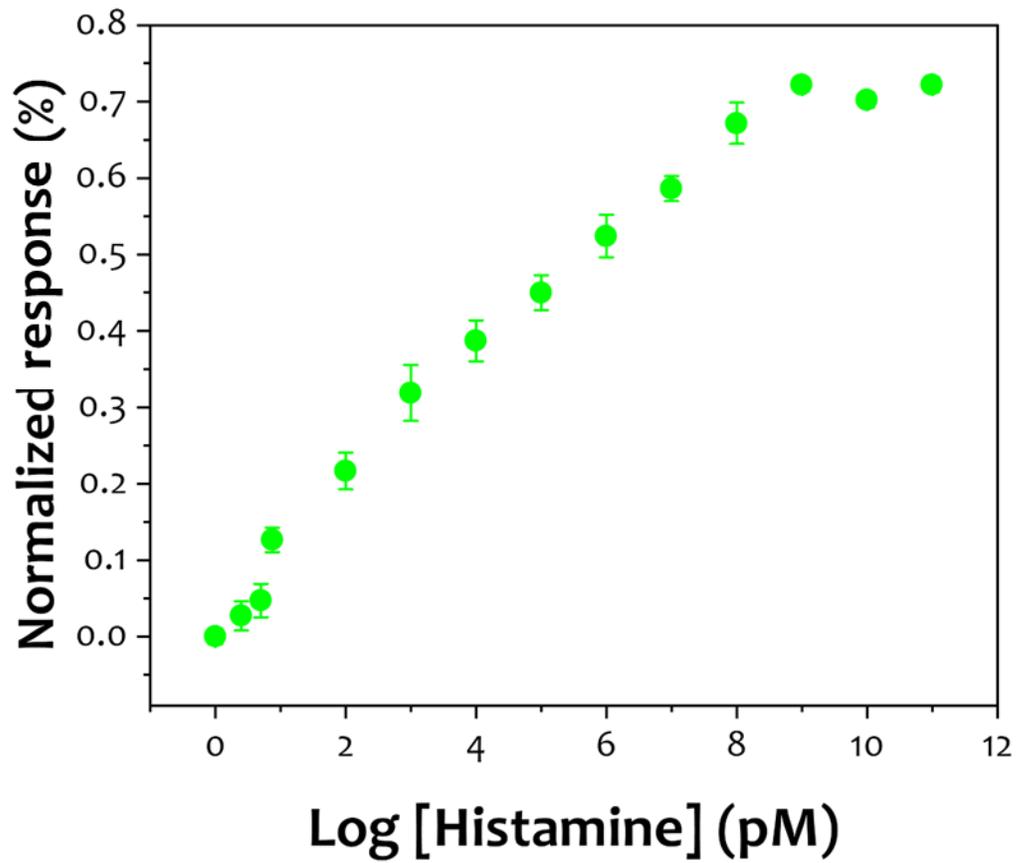

**Figure S15.** Calibration curve for histamine detection in the range of 0.1 pM to 100 mM in 1x PBS. Each point is an average of N = 3 independent sensors.

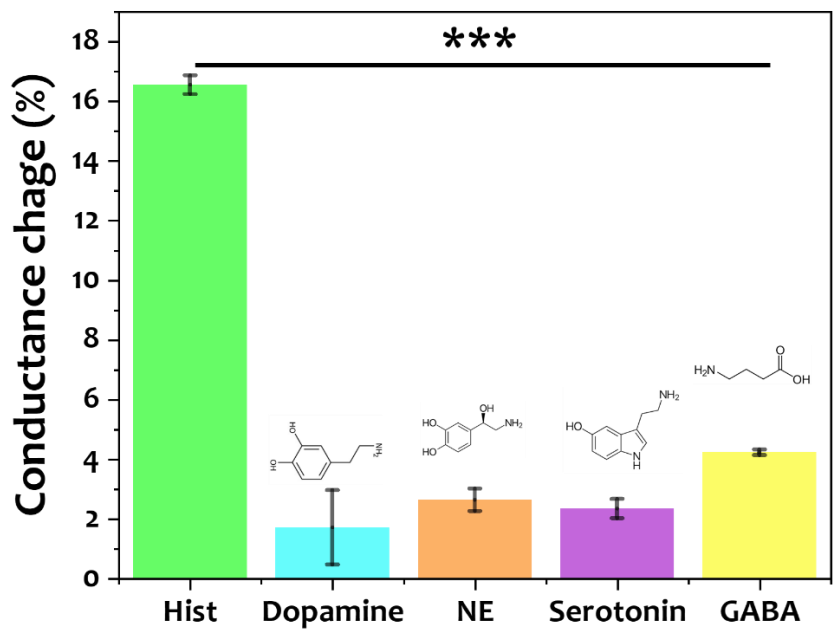

**Figure S16.** Selectivity test of the histamine-MIP nanopore against 1 uM of different analytes, dopamine, GABA, Histamine, NE, and serotonin in neurobasal medium. Each point is an average of $N = 3$ independent sensors. [one-way ANOVA, ***$p < 0.0001$]

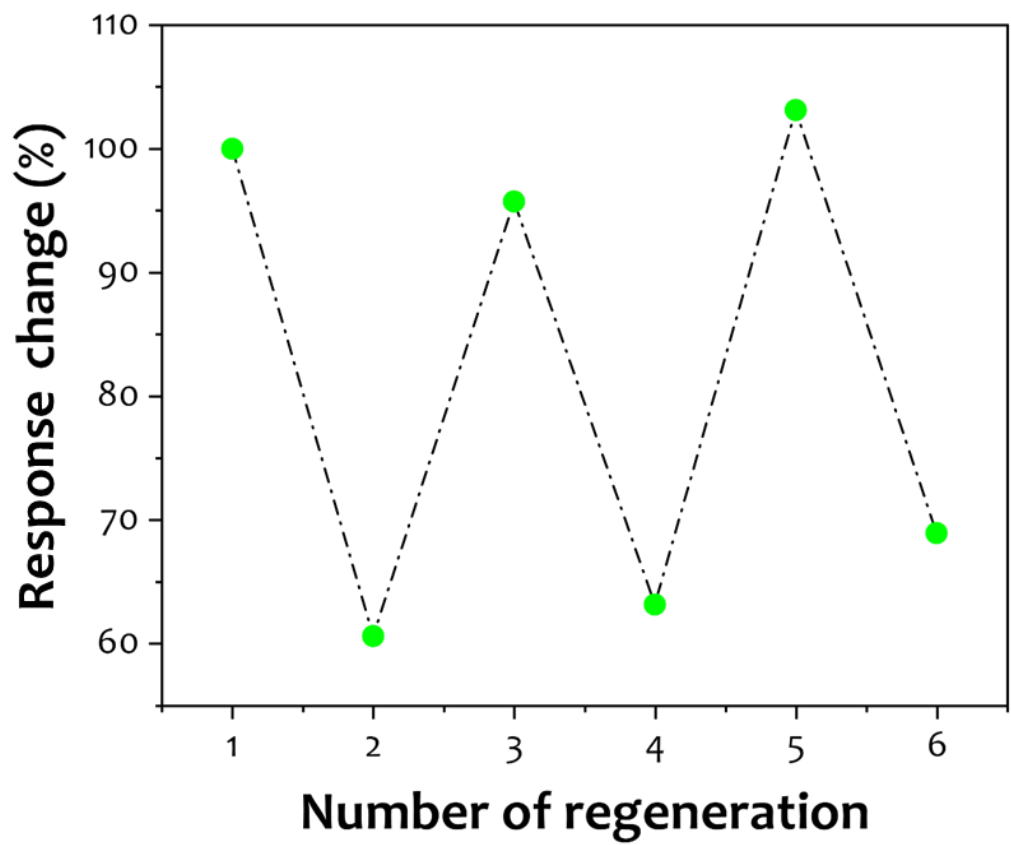

**Figure S17.** The recovery of the His-MIP nanopore in presence of 1 μM histamine.

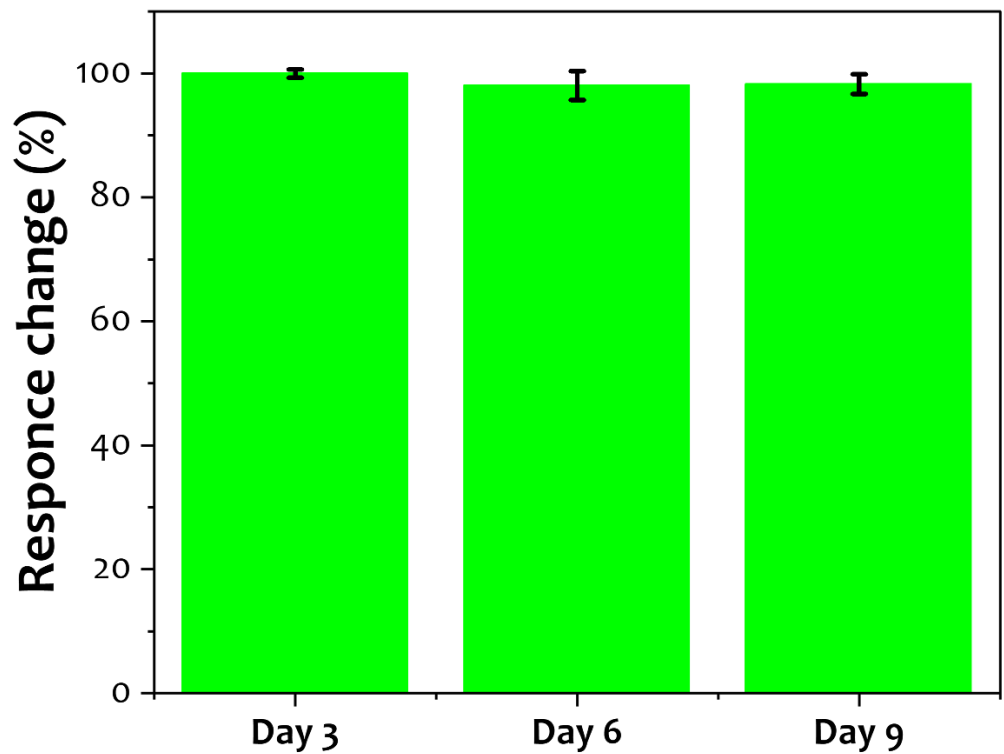

**Figure S18.** Stability test of the His-MIP nanopore for several days. Each point is an average of $N = 3$ independent sensors.

# Note 6. Multiplex sensing and releasing

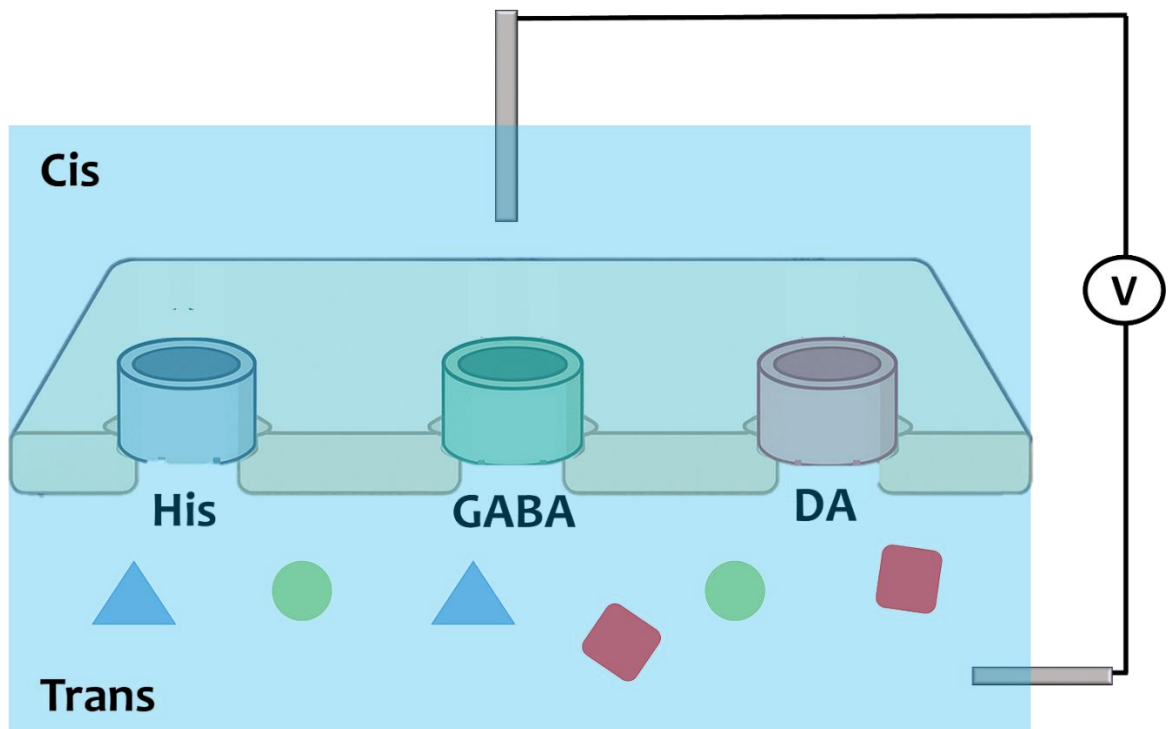

**Figure S19.** Conventional solid-state nanopore electrical measurements illustrate the challenge of distinguishing current changes originating from individual nanopores when they share the same electrolyte chambers as the current change takes into consideration the changes from all the three nanopores.

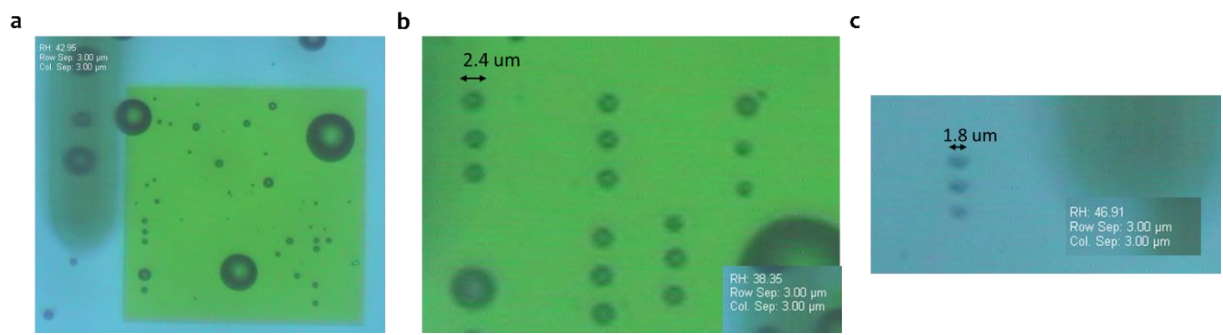

**Figure S20. Effect of humidity on hydrogel micro-spotting with the BioForce Nano eNabler system.** Optical images show hydrogel spots deposited at different relative humidities (RH), with constant row and column spacing of 3 µm. At RH ≈ 43% (a), droplets spread out once in contact with the substrate, therefore, resulting in larger droplets. At RH ≈ 38% (b), spots were more uniform with an average diameter of ~2.4 µm. However, At RH ≈ 47% (c), the optimized deposition conditions yielded well-defined spots with diameters as small as ~1.8 µm. These results highlight the critical role of ambient humidity in controlling hydrogel spot size and the possibility of achieving spots with low spacing.

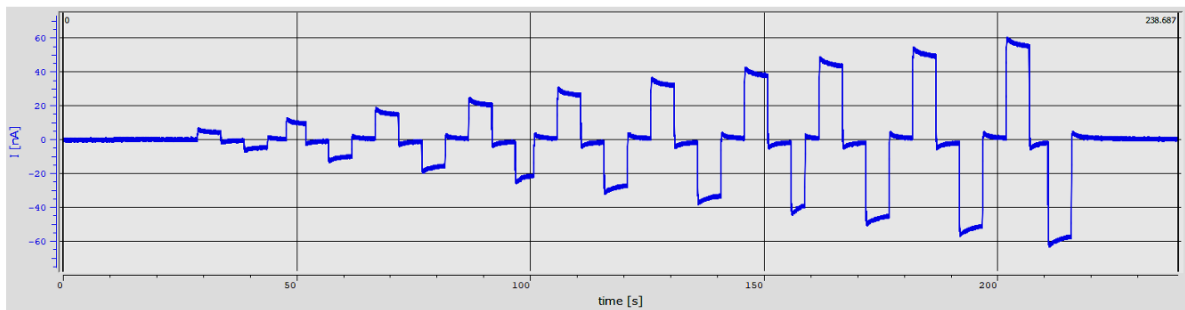
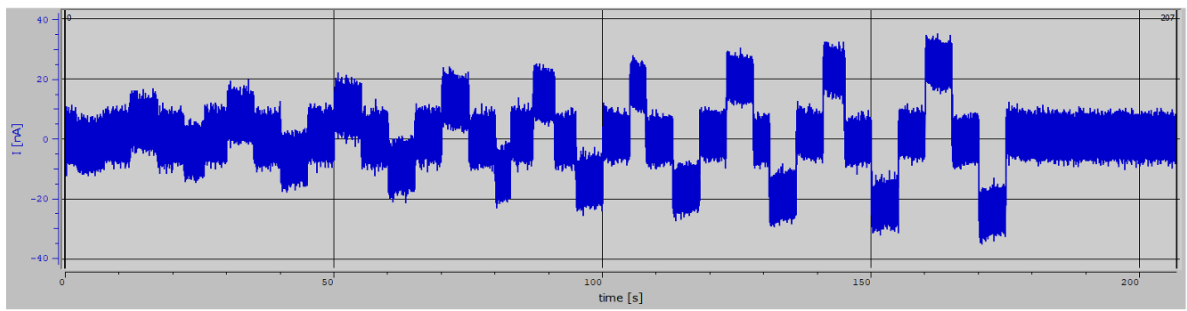
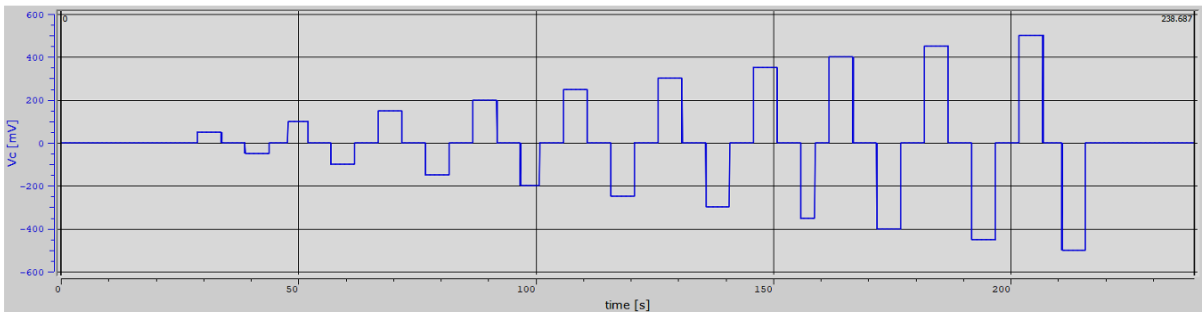

**Figure S21.** a) Ionic current measured from a nanopore over-time at different voltages with 1× PBS in both compartments. b) Ionic current measured from a nanopore with hydrogel in the top chamber and 1× PBS in the bottom chamber. c) The applied voltage over-time.

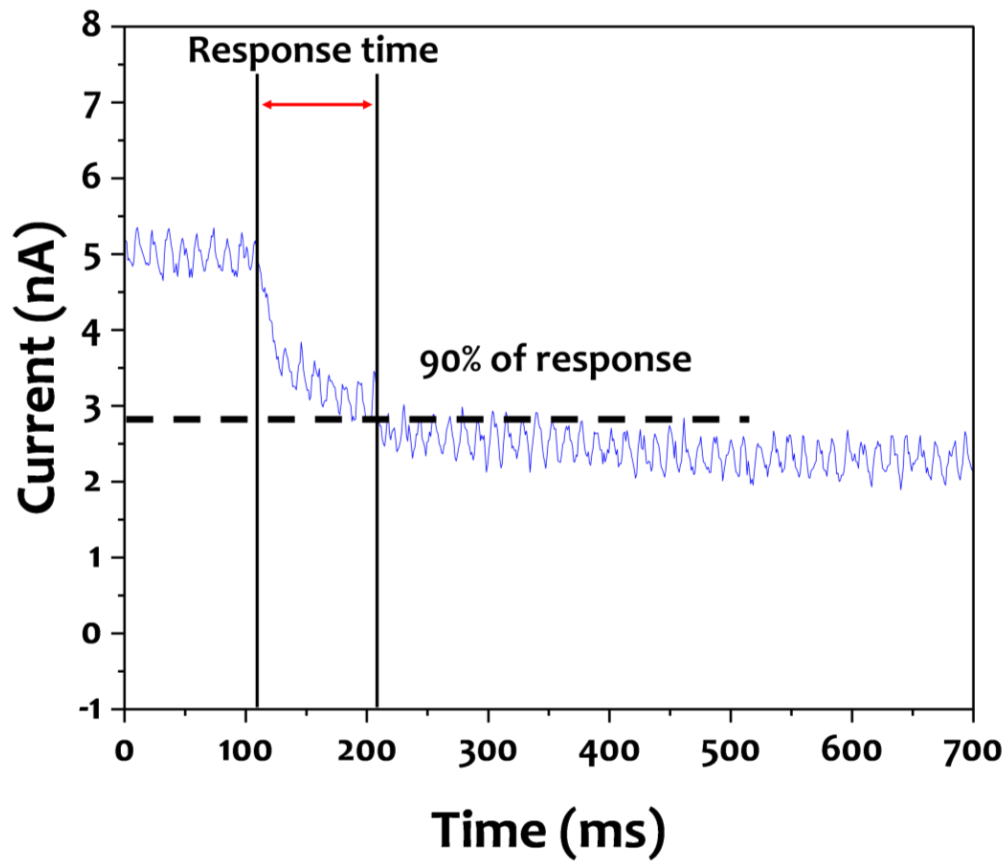

**Figure S22.** The determination of the response time of the dopamine MIP-sensor in neurobasal medium in response to 1 µM of dopamine.

**Note 7. Releasing mechanism**

Figure 3c highlights a key feature of this platform, namely its selective release mechanism, applying an electrical pulse to the DA-MIP nanopore exclusively releases dopamine without causing the release of DOX, and vice versa. Importantly, the long-term retention of bound analytes is inherently linked to the affinity–reversibility trade-off of MIPs. While the high binding affinity of the MIP enables sensitive and selective detection, it may also lead to slow unbinding kinetics in the absence of an external trigger "stimuli", which can limit spontaneous release and extend retention times [3,4]. In our case, the negligible leakage observed over the course of the experiment suggests that unbinding and diffusion are minimal without stimulation, and that active electrical pulses provide a reliable means of release. Several mechanisms have been proposed for molecular release from MIP-based systems, including electrochemical oxidation, local Joule heating, and electrostatic repulsion [5]. In our configuration, electrochemical oxidation can be excluded, as it is not expected to contribute significantly under the applied conditions. We therefore focused on distinguishing between Joule heating and electrostatic effects. To evaluate the potential contribution of heating, we integrated a nanoheater into the nanopore platform (Figure S20) and monitored dopamine release, measured as changes in nanopore conductance, at three different temperatures, room temperature, 40 °C, and 60 °C. No appreciable release was observed at elevated temperatures, ruling out Joule heating as a dominant release pathway. We next investigated the role of applied electric fields through different pulse times. When voltage pulses of increasing duration were applied, we consistently observed rapid change in conductance hence, an increase in dopamine's release (Figure S21). Given the strong electrostatic interactions between the positively charged amine group of dopamine and the negatively charged carboxylates in the polymer matrix, we attribute this process to voltage-induced electrostatic desorption. Once desorbed, dopamine molecules are transported through the nanopore by a combination of electrophoretic and electroosmotic forces, completing the release process.

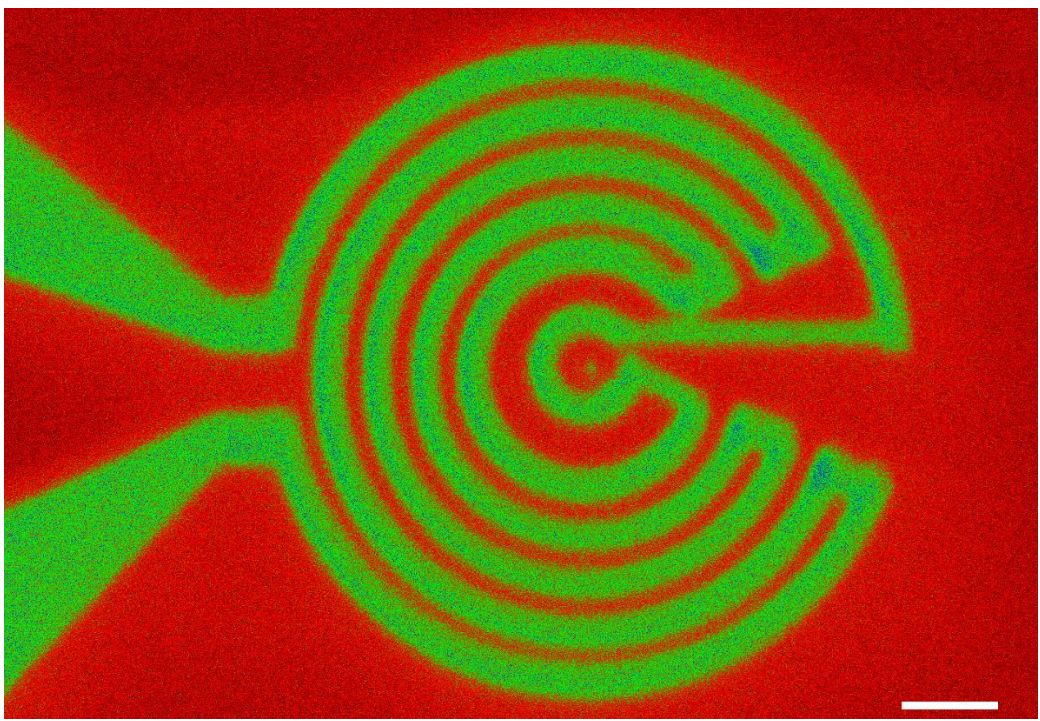

**Figure S23.** A false-colored scanning electron micrograph of a nanoheater-embedded nanopore fabricated in a SiN$_x$ membrane, green color shows the platinum coil (scale bar denotes 500 nm).

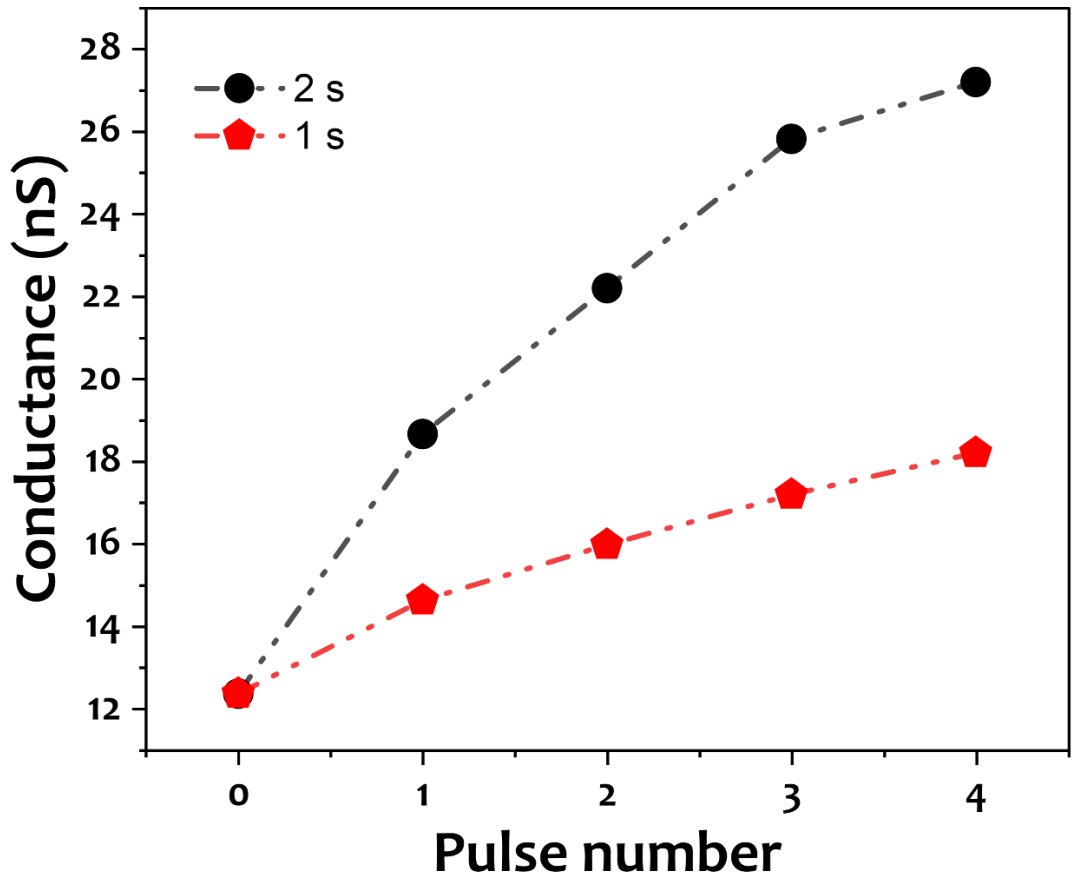

**Figure S24.** Correlation between the pulse time and the change in conductance (dopamine release) in 1x PBS. Showing a correlation between the pulse time and the release, therefore, implying that the primary mechanism is electrostatic repulsion.

## Note 8. Binding affinity of dopamine MIP

To get a deeper understanding of the release mechanism of the MIP nanopore, we first quantified the binding affinity of the MIP by fitting QCM-D kinetics with a 1:1 Langmuir model. In the Sauerbrey regime (rigid, thin adlayer (20 nm measured with AFM); small $\Delta D$), the frequency shift $\Delta f$ is proportional to the surface-bound mass, so $\Delta f(t)$ reports the surface occupancy $\theta(t)$ [6].

- Association (at 10 nM dopamine):

$$\Delta f(t) = \Delta f max\,(1 - e^{-k_{obs}(t-t_0)}),\ with\ k_{obs} = k_a C + k_d$$

Where $k_a$ is the association constant (M$^{-1}$ s$^{-1}$) and $k_d$ the dissociation constant (s$^{-1}$). Moreover, we obtained $k_{obs}$ with nonlinear least squares fitting.

- Dissociation (buffer exchange at t=t$_{switch}$ ):

$$\Delta f(t) = \Delta f(t_{switch})\,e^{-k_d(t-t_{switch})} \qquad (2)$$

From these two fits we calculated the on-rate ka=(k$_{obs}$−k$_d$)/C and the equilibrium dissociation constant K$_D$=k$_d$/ka, with C = 10 nM of dopamine. Applying the fitting to the dataset in Fig. S22 gave:

$$k_d = 2.97 \times 10^{-3}\ s^{-1}$$
$$k_a = 5.36 \times 10\,5\ M - 1s - 1$$

and thus $k_a = 5.54 \times 10^{-9}$ with $k_a C = 5.36\ x\ 10^{-3}\ s^{-1}$ and $k_{obs} = 8.33\ 10^{-3} s^{-1}$. And finally, K$_D$ was calculated to be $KD = 5.54\ x\ 10^{-9}\ M\ (5.5\ nM)$

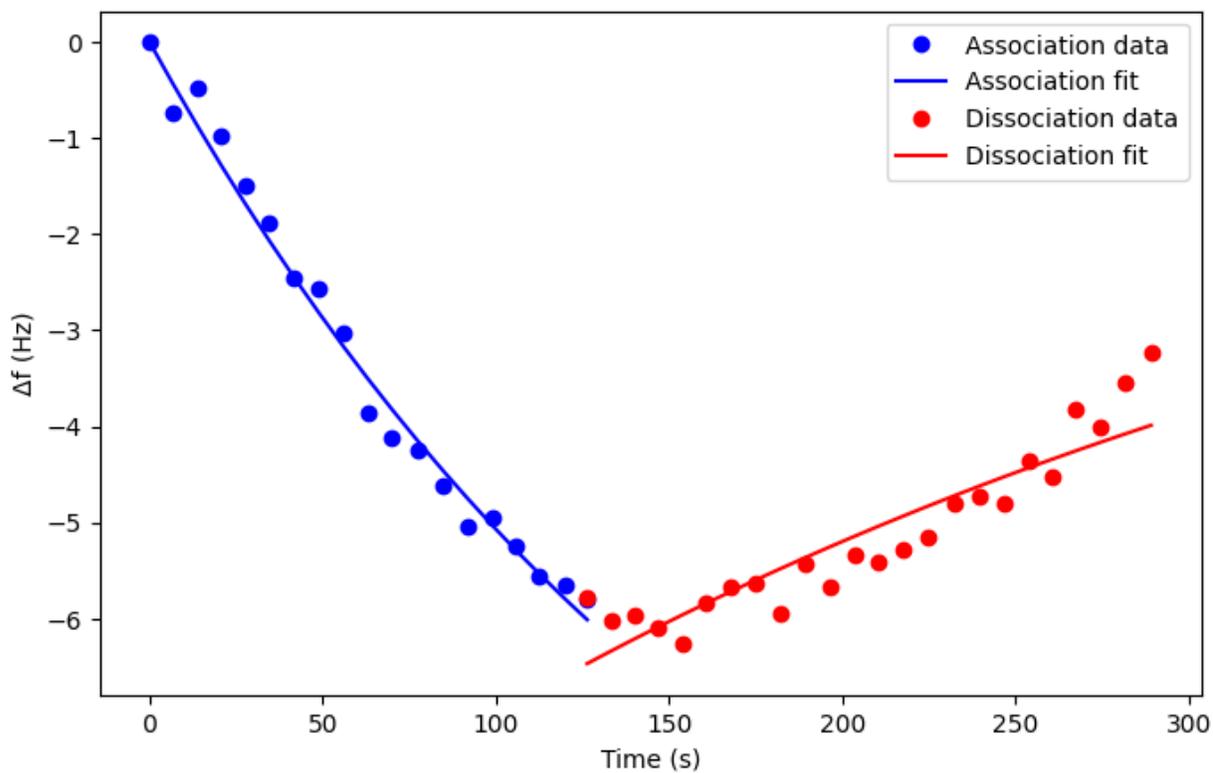

**Figure S25.** QCM-D association and dissociation curves for dopamine binding measured at the 7th harmonic using QCM gold electrode. The frequency shift (Δf) is shown as a function of time. Blue circles represent experimental association data with the corresponding exponential fit (blue line), while red circles represent dissociation data with the corresponding exponential fit (red line).

## Note 9. Field-assisted desorption model, parameterization with pulse-time data, and implications

Once we calculated the binding affinity and probe the rapid release of dopamine observed under voltage pulses, we modeled the desorption kinetics using the framework of Liu et al [7]. Combined with our experimental data from QCM-D (Note. 8) and voltage-pulse release (Fig. S22). This approach allowed us to decouple the thermal desorption rates from field-assisted acceleration and to quantify the effective local potential drop at the binding site.

### a) Field-Assisted Desorption Model

In a nanopore of thickness h under a transmembrane bias U, the average field is $E = \frac{U}{h}$. For a bound ligand with effective valence n (dopamine n ≈ +1), the force exerted by the field is $F = \eta\, e\, E$. Following the Bell-type description adopted by Liu et al., the desorption rate constant under field becomes:

$$k_{off}(V) = k_{off}(0)\, exp\left(\frac{U\, \eta\, e\, z}{h\, k_B\, T}\right)$$

where z is the reactive compliance (distance from bound minimum to transition state, typically 0.1–0.5 nm), $k_B$ is Boltzmann's constant, and T is temperature.

### b) Baseline Kinetics from QCM-D

From QCM-D binding experiments we obtained association and dissociation rates:

$k_a = 5.36\ x\ 10^5 M^{-1} s^{-1}$
$k_{off} = 2.97\ x\ 10^{-3} s^{-1}$
$k_D = 5.5\ nM$

These values correspond to a thermal lifetime $\tau_{off}(0) = \frac{1}{k_{off}(0)} \approx 336\ s$, indicating that without an external field, dopamine remains bound on timescales of several minutes.

### c) Pulse-Time Analysis (Fig. S23)

Afterwards, we analyzed the release under 1 s and 2 s voltage pulses. The probability of release during a pulse is[8]:

$P(\tau) = 1 - e^{-k_{off}(V)\,\tau}$

For small arguments ($k_{off}(V) \ll 1$), $P(\tau) \approx k_{off}(V)\,\tau$, implying linear scaling of release with pulse duration. Figure. S23 shows that, experimentally, increasing the pulse time from 1 s to 2 s approximately doubled the release signal, confirming operation in this linear regime. From the slope of the release vs. pulse time, one can estimate $k_{off}(V) \approx 0.2 - 0.4\ s^{-1}$, corresponding to lifetimes of 2.5–5 s. Thus, the applied field accelerates desorption by roughly two orders of magnitude compared to the thermal baseline.

### d) Barrier Lowering and Local Potential Drop

When a voltage is applied across a nanopore, the total voltage is spread out along the length of the pore. However, the electric field is not uniform. The local potential drop is the actual bias that affects the binding equilibrium of dopamine molecule inside the MIP along its escape path. In the case of dopamine with a +1 charge, the field exerts a directional force along the reaction coordinate, reducing the activation barrier for desorption. This phenomenon, often described by a Bell-type model, results in a field-accelerated off-rate [7]:

$$k_{off}(V) = k_{off}(0) \exp\left(\frac{U \eta e z}{h k_B T}\right)$$

The required acceleration factor is:

$$\frac{k_{off}(V)}{k_{off}(0)} \approx 67\text{--}135$$

Taking the natural logarithm yields:

$$\Delta G_{reduction} \approx 4.2 - 4.9 \, k_B T$$

This corresponds to a local potential drop:

$$\Delta \varphi_{site} \approx \frac{\Delta G_{reduction} \, k_B T}{e} \approx 0.11 - 0.13 \, V \quad \text{(Eq. S3)}$$

Our results match what Liu et al. found. They showed that even the strongest known biomolecular bond, like biotin-avidin (with a dissociation constant around $K_D \sim 10^{-15}$ M), can break within minutes when exposed to electric fields of hundreds of millivolts in nanopores. This reduces how long the bond lasts by four orders of magnitude. In this study, dopamine-MIP interaction speeds up by about 100 times, with a local voltage drop of about 0.1 volts at the binding site. Together, these findings show that the release of molecules from MIP nanopores when voltage is applied is due to field-assisted desorption, not because of heat or Joule heating. Experiments at temperatures between 25 and 60 degrees Celsius ruled out thermal effects. After the molecule comes off, it moves quickly, faster than a millionth of a second, making desorption the main factor that limits the speed. The short current bursts, about 0.1 seconds long (Fig. S22), show the fastest sites, while the longer pulses, lasting 1 to 2 seconds, represent the average behavior of all the sites. So, the voltage lowers the electrostatic barrier by about 4.2 to 4.9 times the thermal energy ($4.2 - 4.9 \, k_B T$), which is roughly 0.1 volts. This gives a clear explanation for how molecules are released from MIP nanopores when voltage is applied.

### e) Electrophoretic and Electroosmotic Transport After Desorption

Once the molecule binding is disrupted by the electric field, the subsequent transport of dopamine through the nanopore is governed by electrophoretic (EP) and electroosmotic (EOF) forces. These forces act concurrently to drive positively charged molecules from the cis reservoir to the trans compartment once they are released from the MIP binding sites.

- **Electrophoretic force**

The positively charged dopamine at physiological pH and under an applied transmembrane field $E = \frac{U}{h}$, it experiences an electrophoretic force $F_{ep} = qE$, resulting in a drift velocity $v_{ep} = \mu E$, where electrophoretic mobility is $\mu$[7]. For small cations, $\mu \approx 2 - 4 * 10^{-8} m^2/Vs$

With U = 1 V across a 50 nm pore, the drift velocity is $v_{ep} \approx 0.6 \, m/s$, corresponding to a transit time of ~80 ns across the pore.

- **Electroosmotic flow**

The carboxylated MIP matrix and $SiN_x$ pore walls carry negative charges, producing an electrical double layer [7]. When an electric field is applied, the counterion layer drags solvent, generating EOF. The EOF velocity is described by Smoluchowski's relation[9]:

$$v_{EOF} \approx -\left(\frac{\varepsilon \zeta}{\eta}\right) E$$

where $\varepsilon$ is permittivity, $\zeta$ is zeta potential, and $\eta$ is viscosity. For water ($\varepsilon \approx 7 \times 10^{-10}$ C/Vm, $\eta \approx 10^{-3}$ Pa·s) and $\zeta \approx$ -30 mV. Hence, the EOF velocity at 1V and 50 nm is $v_{EOF} \approx 0.4 \, m/s$ .

- **Combined effect and implications**

Now, as the effect of both EP and EOF act in the same direction under positive bias results in combined effect exerted on the dopamine molecules. The combined transport velocities (~1 m/s) yield transit times below 1 µs for a 50 nm pore. This is orders of magnitude faster than the observed release dynamics "0.1 s", this implies that the post-desorption transport is not rate limiting. Instead, the rate-limiting step is field-assisted electrostatic desorption at the MIP binding cavities.

**Note 10. Ionic neurotransmitter-based logic gates**

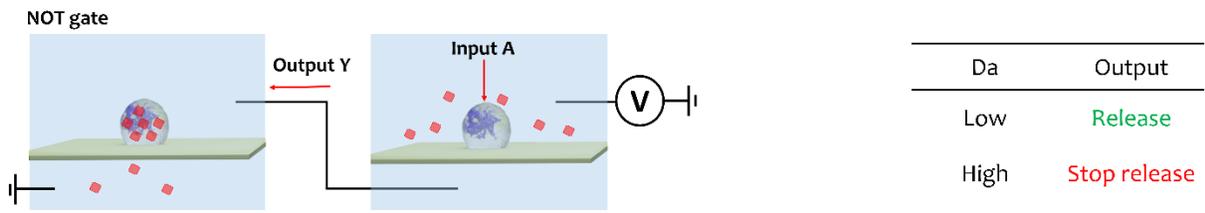

**Figure S26.** Schematic of NOT gate with a truth table used for edge-computing and in-memory sensing.

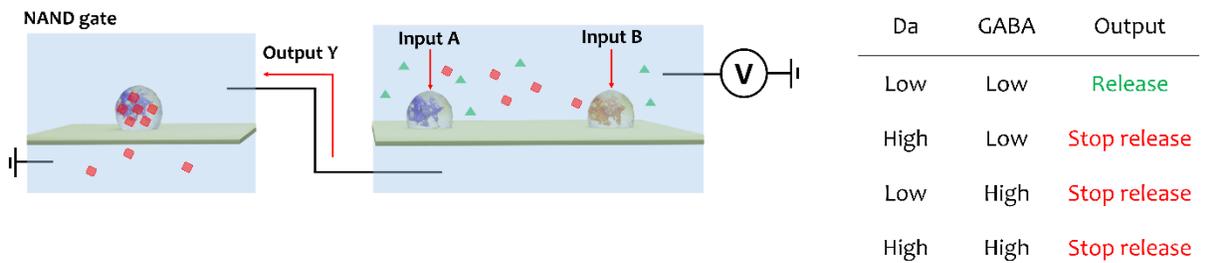

**Figure S27.** Schematic of NAND gate with a truth table used for edge-computing and in-memory sensing.

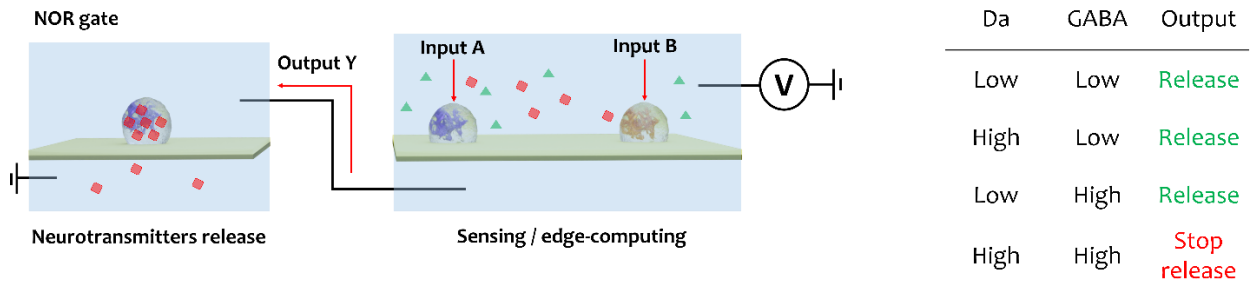

**Figure S28.** Schematic of NOR gate with a truth table used for edge-computing and in-memory sensing.